\documentclass[fleqn,usenatbib]{mnras}
\usepackage{amsmath,amssymb,pgf,graphicx,textpos,color}
\usepackage{hyperref}

\newcommand{\be}{\begin{equation}}
\newcommand{\ee}{\end{equation}}

\newcommand{\black}{\color{black}}

\title[Point sources of UHE photons with TA SD]{Search for point sources
of ultra-high energy photons with the Telescope Array surface detector}

\author[Telescope Array Collaboration]{
  \parbox{\textwidth}{
Telescope Array Collaboration:
R.U.~Abbasi$^{1}$,
M.~Abe$^{2}$,
T.~Abu-Zayyad$^{1}$,
M.~Allen$^{1}$,
R.~Azuma$^{3}$,
E.~Barcikowski$^{1}$,
J.W.~Belz$^{1}$,
D.R.~Bergman$^{1}$,
S.A.~Blake$^{1}$,
R.~Cady$^{1}$
% ...
\\
(Affiliations can be found after the references)
}}

\begin{document}

\date{Accepted . Received ; in original form }

\pagerange{\pageref{firstpage}--\pageref{lastpage}} \pubyear{2015}

\maketitle
\clearpage
\label{firstpage}
\begin{abstract}
The surface detector (SD) of the Telescope Array (TA) experiment allows one to indirectly detect photons 
with energies of order $10^{18}$~eV and higher and to separate photons from the cosmic-ray background. 
In this paper we present the results of a blind search for point sources
of ultra-high energy (UHE) photons in the Northern sky using the TA SD data.
The photon-induced extensive air showers (EAS) are
separated from the hadron-induced EAS background by means of a multivariate classifier based upon
16 parameters that characterize the air shower events. No significant evidence for the photon point sources
is found. The upper limits are set on the flux of photons  from each particular direction in the sky
within the TA field of view, according to the experiment's angular resolution for photons.
Average 95\% C.L. upper limits for the point-source flux of photons with energies
greater than $10^{18}$, $10^{18.5}$, $10^{19}$, $10^{19.5}$ and $10^{20}$~eV
are $0.094$, $0.029$, $0.010$, $0.0073$ and $0.0058$~km$^{-2}$yr$^{-1}$ respectively.
For the energies higher than $10^{18.5}$~eV, the photon point-source limits are set
for the first time. Numerical results for each given direction in each energy range are
 provided as a supplement to this paper. 
\end{abstract}

\section{Introduction}
Ultra-high energy photons are an important tool for studying the high-energy Universe.
A plausible source of EeV-energy photons is provided by ultra-high energy cosmic rays (UHECR) undergoing the
Greisen-Zatsepin-Kuzmin process~\citep{Greisen:1966jv, Zatsepin:1966jv} or pair production process~\citep{Blumenthal:1970nn} 
on a cosmic background radiation. In this context, the EeV photons can be a probe of UHECR mass composition
as well as of the distribution of their sources~\citep{Gelmini:2005wu, Hooper:2010ze}.
At the same time, the possible flux of photons produced by UHE protons
in the vicinity of their sources by pion photoproduction or inelastic nuclear collisions
would be noticeable only for relatively near sources, as the UHE photons attenuation
length is smaller than that of UHE protons (see e.g.~\cite{Bhattacharjee:1998qc} for a review). 
There also exists a class of so-called top-down models of UHECR generation that efficiently produce the UHE photons,
for instance by the decay of heavy dark-matter particles~\citep{Berezinsky:1997hy, Kuzmin:1997jua}
or by the radiation from cosmic strings~\citep{Berezinsky:1998ft}. The search for the UHE photons was shown to be the most sensitive
method of indirect detection of heavy dark matter~\citep{Kalashev:2016cre, Kalashev:2017ijd, Kuznetsov:2016fjt, Kachelriess:2018rty, 2019arXiv190305429A}. 
Another fundamental physics scenario that could be tested with UHE photons~\citep{Fairbairn:2009zi}
is the photon mixing with the axion-like particles~\citep{Raffelt:1987im} that could be responsible for the correlation of
UHECR events with BL Lac type objects observed by the HiRes experiment~\citep{Gorbunov:2004bs, Abbasi:2005qy}.
In most of these scenarios, clustering of photon arrival directions rather than diffuse distribution
is expected, therefore point-source searches can be a suitable test for them.
Finally, the UHE photons could also be used as a probe for the models of Lorentz-invariance
violation~\citep{Coleman:1998ti, Galaverni:2007tq, Maccione:2010sv, Rubtsov:2012kb, Rubtsov:2013wwa}.

Telescope Array~\citep{AbuZayyad:2012kk, Tokuno:2012mi} is the largest cosmic-ray experiment in the Northern Hemisphere.
It is located at $39.3^\circ$~N, $112.9^\circ$~W in Utah, USA. 
The observatory includes a surface detector array (SD) and 38 fluorescence telescopes grouped into three stations.
The SD consists of 507 stations that contain plastic scintillators each of 3 m$^2$ area (SD stations).
The stations are placed in the square grid with the 1.2 km spacing and covers the area of $\sim 700 \; {\rm km}^2$.
The TA SD is capable of detecting EAS in the atmosphere caused by cosmic particles of EeV and higher energies.
The TA SD operates since May 2008. 

A hadron-induced extensive air shower (EAS) significantly differs from an EAS induced by a photon:
the depth of the shower maximum $X_{\rm max}$ for a photon shower is larger, a photon shower
 contains less muons and has more curved front  (see~\citep{Risse:2007sd} for the review).
The TA SD stations are sensitive to both muon and electromagnetic component of the shower
and therefore may be triggered by both hadron-induced and photon-induced EAS. 

In the present study, we use 9 years of TA SD data for a blind search for
point sources of UHE photons. We utilize the
statistics of the SD data, which benefits from high duty cycle. 
The full Monte-Carlo (MC) simulation of proton-induced and photon-induces EAS events allows us to perform the
photon search up to the highest accessible energies, $E \gtrsim 10^{20}$~eV. As the main tool for the
present photon search we use a multivariate analysis based on a number of the SD parameters 
that make possible to distinguish between photon and hadron primaries. 

While searches for diffuse UHE photons were performed by several EAS experiments,
including Haverah Park~\citep{Ave:2000nd}, AGASA~\citep{Shinozaki:2002ve, Risse:2005jr},
Yakutsk~\citep{Glushkov:2007ss, Glushkov:2009tn, Rubtsov:2006tt},
Pierre Auger~\citep{Abraham:2006ar, Aglietta:2007yx, Bleve:2015nut, Aab:2016agp} and
TA~\citep{Abu-Zayyad:2013dii, Abbasi:2018ywn}, the search for point sources of
photons at ultra-high energies has been  done  only by the Pierre Auger Observatory~\citep{Aab:2014bha, Aab:2016bpi}.
The latter searches were based on the hybrid data and were limited to $10^{17.3} < E < 10^{18.5}$~eV
energy range. In the present paper we use the TA SD data alone.
We perform the searches in five energy ranges, namely
$E>10^{18}$, $E>10^{18.5}$, $E>10^{19}$, $E>10^{19.5}$ and $E>10^{20}$~eV.
 We find \black no significant evidence of photon point sources
in all energy ranges and set the point-source flux upper limits from each direction in TA field of view. 
The search for unspecified neutral particles was also previously performed by
the Telescope Array~\citep{Abbasi:2014wza}. The limit on the the neutral particles
point-source flux obtained in that work is close to the present
photon point-source flux limits.

\section{TA SD data and reconstruction}
\subsection{Data set and Monte-Carlo}
\label{data-mc}
The data and Monte-Carlo sets used in this study are the same as in
the recent TA search for diffuse photons~\citep{Abbasi:2018ywn}.
We use the TA SD data set obtained
in 9 years of observation, from May 11, 2008 to May 10, 2017. 
During this period, the duty cycle of the SD was about 95\%~\citep{AbuZayyad:2012ru, Matthews:2017waf}. 

Monte-Carlo simulations used in this study reproduce 9 years of TA SD observations, as it was shown in~\citep{Matthews:2017waf}.
We simulate separately showers induced by photon and proton primaries for the signal and background estimation
respectively$^{1)}$\let\thefootnote\relax\footnote{$^{1)}$ We justify the proton background assumption in the Sec.~\ref{mva}.},
using the CORSIKA code~\citep{Heck:1998vt}. The high energy nuclear interactions are simulated with
QGSJET-II-03 model~\citep{Ostapchenko:2004ss}, 
the low energy nuclear reactions with FLUKA package~\citep{Ferrari:2005zk} and the electromagnetic shower component
with EGS4 model~\citep{Nelson:1985ec}. The usage of the PRESHOWER package~\citep{Homola:2003ru}
that takes into account the splitting of the UHE photon primaries into the Earth's
magnetic field allows us to correctly simulate photon-induced EAS up to the
100 EeV primary energy and higher. The thinning and dethinnig procedures with 
parameters described in~\citep{Stokes:2011wf} are used to reduce the calculation time.

We simulated 2100 CORSIKA showers for photon
primaries and 9800 for proton primaries in $10^{17.5}-10^{20.5}$~eV primary energy range.
The power spectrum for CORSIKA photon events is set to $E^{-1}$. 
The showers from the photon and the proton libraries are processed by the code simulating
the real time calibration SD response by means of GEANT4 package~\citep{Agostinelli:2002hh}.
Each CORSIKA event is thrown to the random locations within the SD area multiple times. 
 For photons, these procedures also include reweighting of the events to the $E^{-2}$ differential spectrum,
 which is assumed for primary photons in this work. 
As a result, a set of 57 million photon events with $E^{-2}$ spectrum was obtained.
The proton Monte-Carlo set used in this study contains approximately 210 million of events.
 Details of proton Monte-Carlo simulations are described in Refs.~\citep{AbuZayyad:2012ru, 2014arXiv1403.0644T, Matthews:2017waf}. 
The format of the Monte-Carlo events is the same as the one used for real events, therefore both
data and Monte-Carlo are processed by one and the same reconstruction procedure~\citep{2014arXiv1403.0644T} described below.

\subsection{Reconstruction}
\label{rec}
In this paper, the same procedure to reconstruct shower parameters is used
as in the previous TA photon  searches~\citep{Abu-Zayyad:2013dii, Abbasi:2018ywn}. 
 Each event real or simulated,  is reconstructed by a joint fit of the shower-front 
geometry and the lateral distribution function (LDF) that allows us to determine the shower parameters, including the 
arrival direction, the core location, the signal density at the fixed distance from the core
and the shower front curvature parameter (see~\citep{Abu-Zayyad:2013dii} for details). 

We apply the following set of the quality cuts for both MC and data events:
\begin{enumerate}
\item zenith angle cut: $0^\circ < \theta < 60^\circ\,$,
\item the number of stations triggered is 7 or more,
\item the shower core is inside the array boundary with the
distance to the boundary larger than 1200 m,
\item joint fit quality cut: $\chi^2/{\rm d.o.f.} < 5$.
\end{enumerate}
We also use an additional cut to eliminate the events induced by lightnings.
It was previously  found  by the TA collaboration that lightning strikes could cause
events mimicking EAS events, the so-called terrestrial
gamma-ray flashes (TGF)~\citep{ABBASI20172565, Abbasi:2017muv}.
Moreover, as the lightning events are expected to be electromagnetic, they resemble
photon-induced showers. Therefore, the rejection of these events is crucial
for photon search. To make this rejection, we use the Vaisala lightning database from the U.S.
National Lightning Detection Network (NLDN)~\citep{NLDN1, NLDN2, NLDNurl}.
From this database we extract the list of the NLDN lightning events detected within
a 15--mile radius circle from the Central Laser
Facility of the TA, that contains all the TA SD stations, in a time range from 2008-05-11 to 2017-05-10.
The list contains 31622 events grouped in time in such a way that
a total of 910 astronomical hours contain one or more
lightnings. To clean up all possible lighting-induced events from the data set
we remove all the events that occur within 10 minutes time intervals before or
after the NLDN lightnings. This cut removes the events known to be related
to the TGFs reducing the total exposure only by 0.66\% and the total number
of data events by 0.77\%. 

The basic  observables such as zenith angle,  calculated in the reconstruction procedure 
together with several additional parameters (see below), 
are used to distinguish photon and proton events by means of a multivariate analysis.
Some of the observables are utilizing the features of the experiment's
SD technical design, such as the double-layered scintillators.
The detailed description of these technical parameters is given in~\citep{AbuZayyad:2012kk}. 
The full list of 16 parameters used in the present photon search is
the same as in the TA SD search for diffuse photons~\citep{Abbasi:2018ywn}
and the TA SD composition study~\citep{Abbasi:2018wlq}.  These parameters  are:
\begin{enumerate}
\item Zenith angle, $\theta$.
\item Signal density at 800 m from the shower core, $S_{800}$. 
\item Linsley front curvature parameter, $a$ obtained front the fit of
the shower front with the AGASA-modified Linsley time delay function~\citep{Teshima:1986rq, Abu-Zayyad:2013dii}. 
\item Area-over-peak (AoP) of the signal at 1200 m~\citep{Abraham:2007rj}.
\item AoP slope parameter~\citep{Rubtsov:2015wba}.
\item Number of stations with Level-0 trigger~\citep{AbuZayyad:2012kk} (triggered stations). 
\item Number of stations excluded from the fit of the shower front due to large contribution to $\chi^2$. 
\item $\chi^2/d.o.f.$ of the shower front fit. 
\item $S_b$ parameter for $b=3$; $S_b$ is defined as $b$-th moment of the LDF:
\be
S_b = \sum\limits_{i} \left[S_i \times \left(r_i/r_0\right)^b\right]\,,
\ee
where $S_i$ is the signal of $i$-th station, $r_i$ is the distance
from the shower core to a given station, $r_0=1000$\,m. The sum is
calculated over all triggered non-saturated stations.
The $S_b$ is proposed as a composition-sensitive parameter in~\citep{Ros:2013lxa}. 
\item $S_b$ parameter for $b=4.5$.
\item The sum of signals of all triggered stations of the event. 
\item An average asymmetry of signal at upper and lower layers of the
stations defined as:
\be
\mathcal{A} = \frac{\sum\limits_{i,\alpha}
 |s^{upper}_{i,\alpha}-s^{lower}_{i,\alpha}|}{\sum\limits_{i,\alpha}
 |s^{upper}_{i,\alpha}+s^{lower}_{i,\alpha}|}\,,
\ee
where $s^{upper|lower}_{i,\alpha}$ is the FADC value of upper or lower
layer of $i$-th station at $\alpha$-th time bin. The sum is calculated
over all triggered non-saturated stations over all time bins of the
corresponding FADC traces. 
\item Total number of peaks of FADC trace summed over upper and
lower layers of all triggered stations of the event.
To suppress accidental peaks as a result of FADC noise, we define a peak as a
time bin with a signal above 0.2 Vertical equivalent muons (VEM)
which is higher than a signal of the 3 preceding and 3 consequent time bins. 
\item Number of peaks for the station with the largest signal.
\item Total number of peaks present in the upper layer and not in the lower one, summed over
all triggered stations of the event. 
\item Total number of peaks present in the lower layer and not in the upper one, summed over
all triggered stations of the event. 
\end{enumerate}

\begin{table}
\begin{center}
\begin{tabular}{|c|c|c|}
\hline
$E_\gamma$, eV & $\langle \theta_{\rm rec.} - \theta_{\rm true} \rangle$ & ang. resolution \\
\hline
$> 10^{18.0}$ & $-2.25^\circ$ & $3.00^\circ$ \\ \hline
$> 10^{18.5}$ & $-2.24^\circ$ & $2.92^\circ$ \\ \hline
$> 10^{19.0}$ & $-2.16^\circ$ & $2.64^\circ$ \\ \hline
$> 10^{19.5}$ & $-2.06^\circ$ & $2.21^\circ$ \\ \hline
$> 10^{20.0}$ & $-1.72^\circ$ & $2.06^\circ$ \\ \hline
\end{tabular}
\end{center}
\caption{
Bias in the reconstruction of the zenith angle and angular resolution for the
photon primaries at various energies.
}
\label{gamma_angles}
\end{table} 

For each MC and data event we also define the ``photon energy''
parameter $E_\gamma$ which is the expected energy of the primary particle assuming it is a photon. 
This energy parameter is calculated as the function of the zenith angle and the $S_{800}$ parameter
from the photon MC simulations~\citep{Abu-Zayyad:2013dii}. For proton MC events, as well as for 
the majority of data events, the $E_\gamma$ parameter is not the actual primary energy but merely
a parameter needed for the consistent comparison of proton events 
and possible photon events. It is important to note that for majority
of proton-induced events, the reconstructed $E_\gamma$ parameter
is systematically higher than that of photon-induced events of the
same primary Monte-Carlo energy. For instance, at $\sim 10$~EeV Monte-Carlo energy the mean $E_\gamma$ for
protons is $\sim 40\%$ higher than that for photons, if we assume the averaging over zenith angle. 
Due to this fact the proton background for SD photon search is higher with respect to the
 hypothetical ideal situation when the energy reconstruction bias is independent
of the primary particle type. All the energy values considered in this work is assumed to be $E_\gamma$ values,
unless the other meaning is specified. 

The reconstructed values of shower zenith angle, $\theta_{\rm rec}$, 
for photon primaries are systematically underestimated. 
The possible reason for this is the azimuthal asymmetry of the shower front,
that originates from the fact that the shower arrives younger to the front-side stations 
and older to the back-side ones. The reconstruction bias is defined as a
deviation of the event $\theta_{\rm rec}$ from a real Monte-Carlo zenith angle of this event, $\theta_{\rm true}$. 
The average values of this bias for various energies $E_\gamma$ are given in Table~\ref{gamma_angles}.
In this study we correct both proton and photon Monte-Carlo events and data events
by these average bias values. This correction allows us to restore
arrival directions of possible photon-induced events more accurately, while not affecting
the background of hadron-induced events, which is known to be highly isotropic~\citep{Deligny:2017wbx}. 
Another crucial parameter for the point-source search is the angular resolution
of the experiment. It is defined as a $0.68$ percentile of a distribution 
of Monte-Carlo events over opening angle between event reconstructed arrival direction
and real Monte-Carlo arrival direction. 
The angular resolution of TA SD for proton primaries at ``proton energy'', $E_{\rm p} = 10^{19}$~eV, 
 was estimated to be $1.5^\circ$~\citep{AbuZayyad:2012hv}. 
As it was mentioned above, in the present study we use the reconstruction of~\citep{Abu-Zayyad:2013dii}
for both data and Monte-Carlo events. Using the photon Monte-Carlo set, 
 after applying the zenith-angle bias correction described above, we estimate the angular resolution
for photon primaries at various energies $E_\gamma$. The results are shown in Table~\ref{gamma_angles}.

\section{Analysis}
\subsection{Multivariate analysis}
\label{mva}

\begin{figure*}
\begin{center}
 \includegraphics[width=0.9\columnwidth]{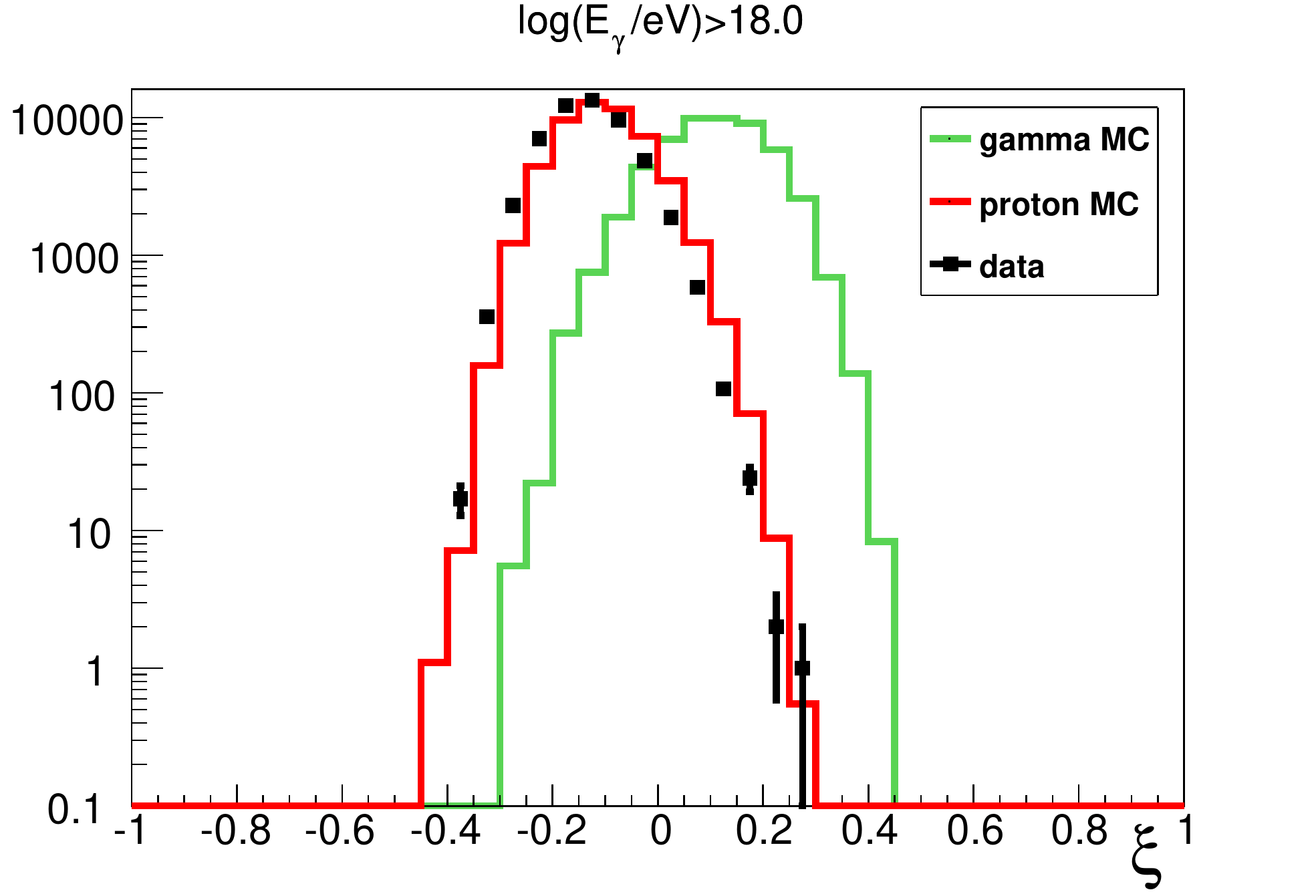}
 \includegraphics[width=0.9\columnwidth]{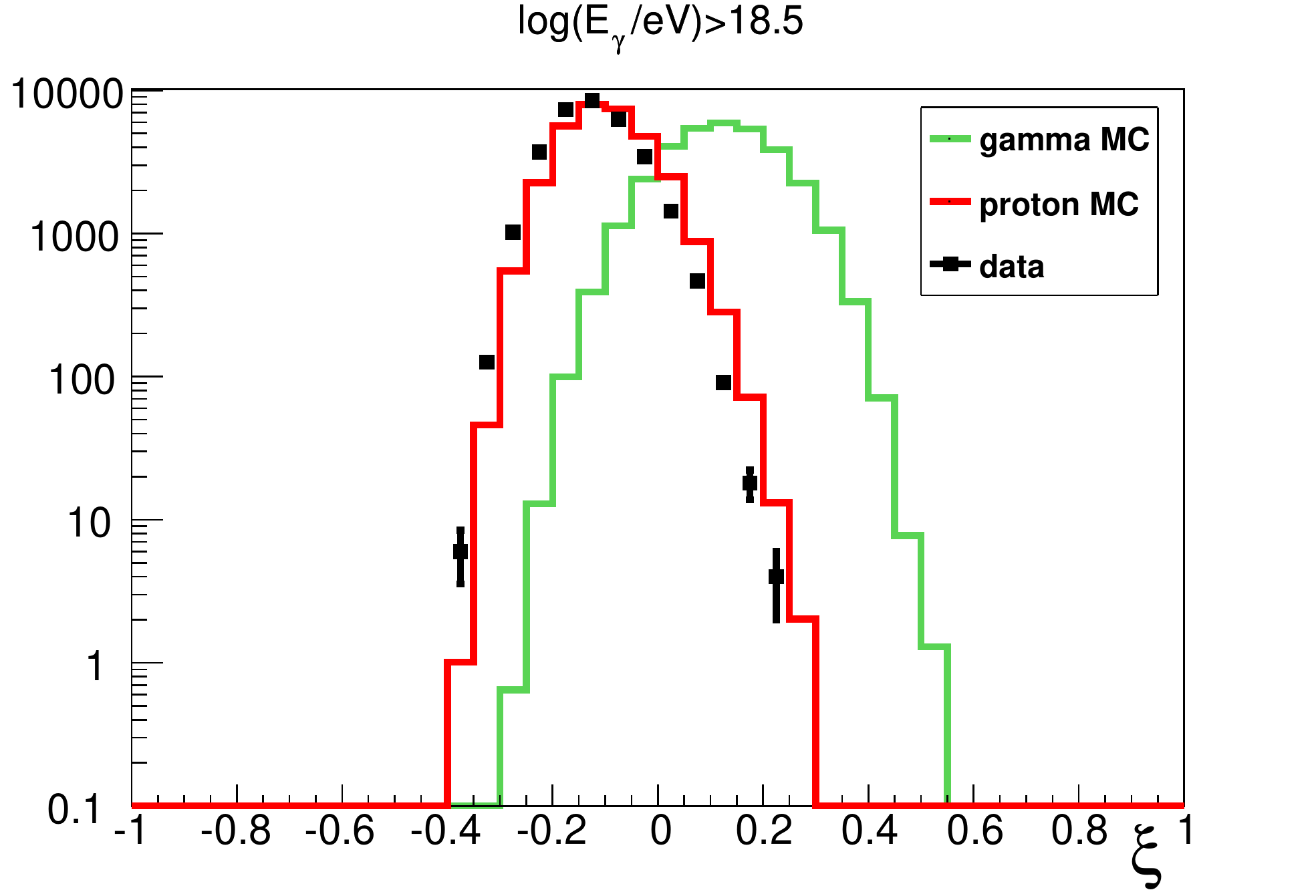}\\
 \includegraphics[width=0.9\columnwidth]{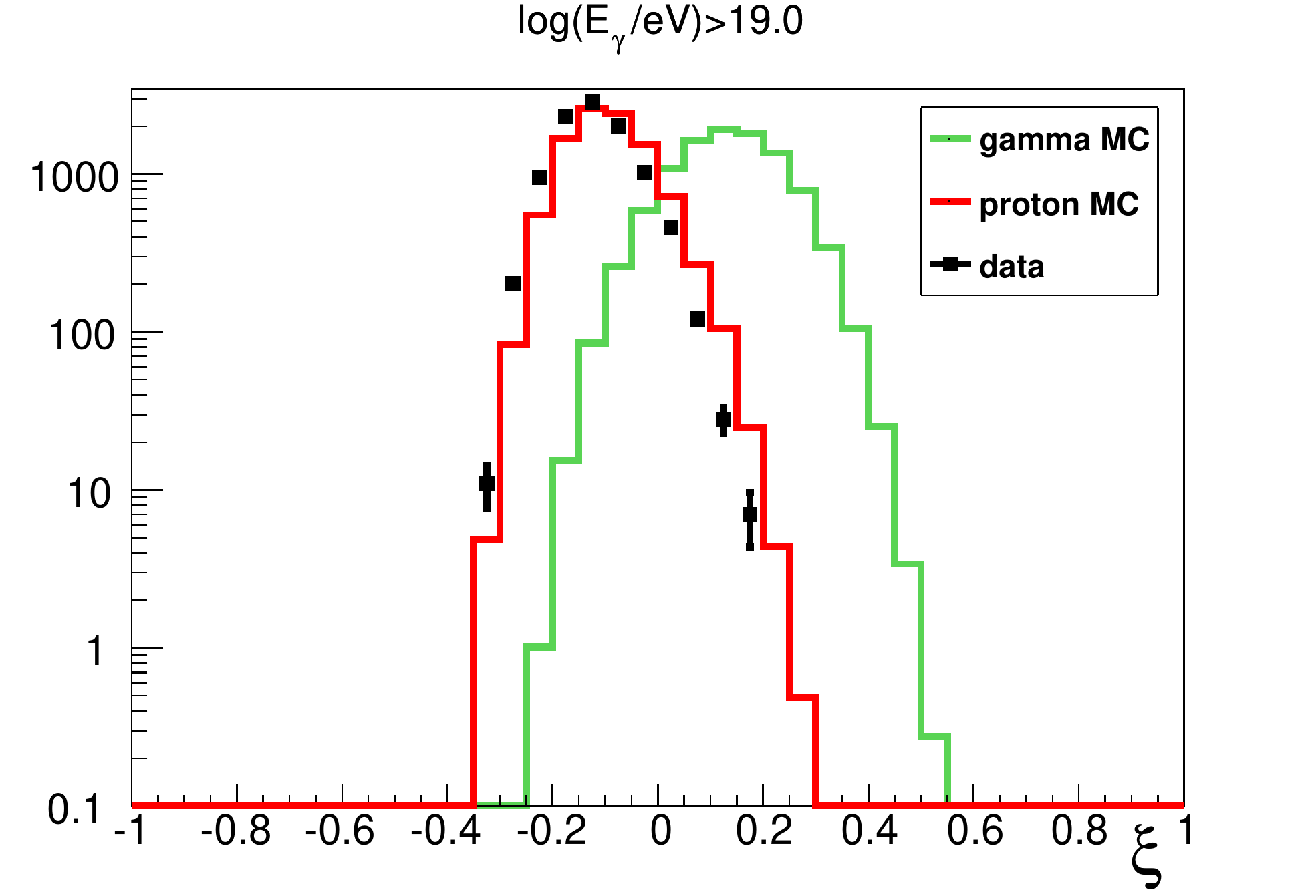}
 \includegraphics[width=0.9\columnwidth]{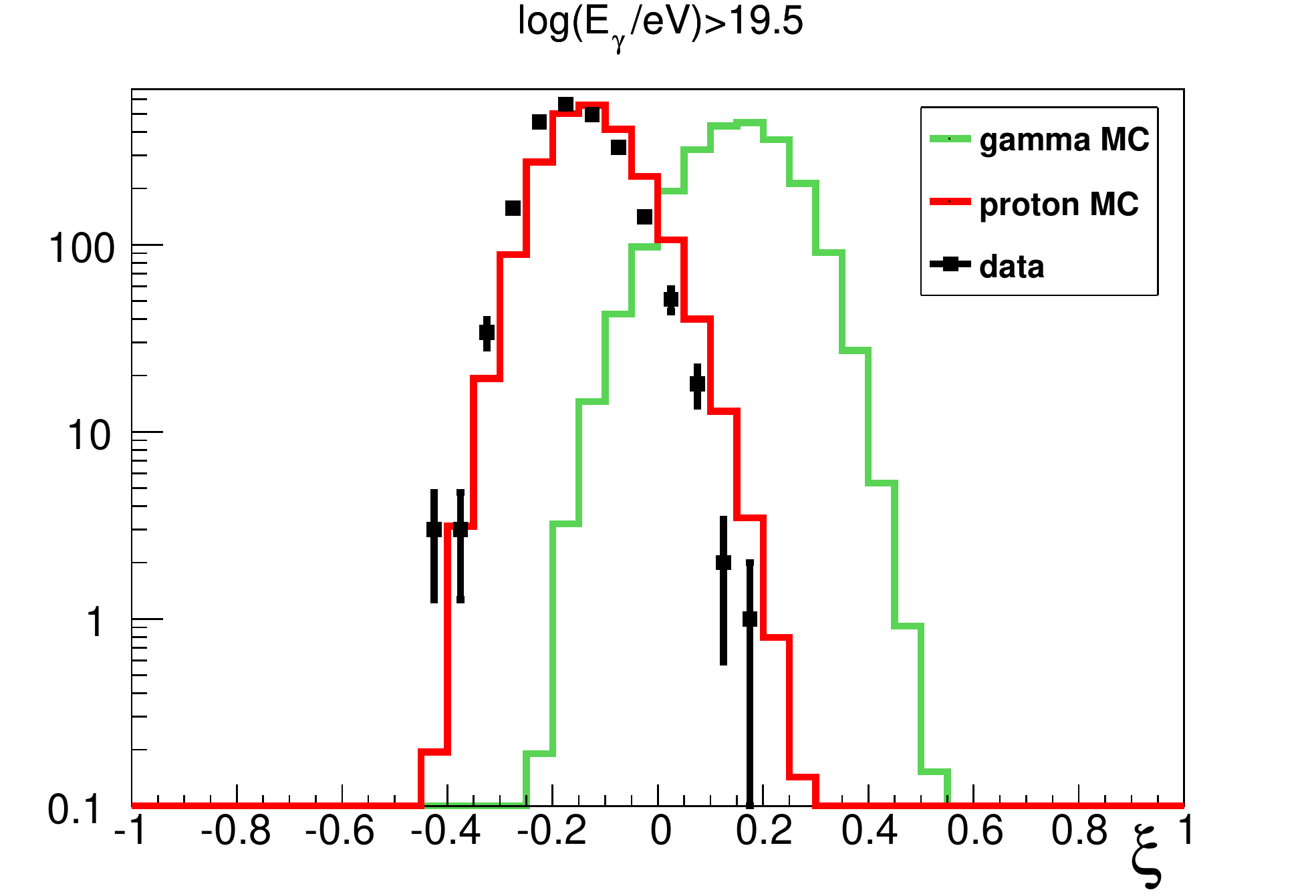}\\
 \includegraphics[width=0.9\columnwidth]{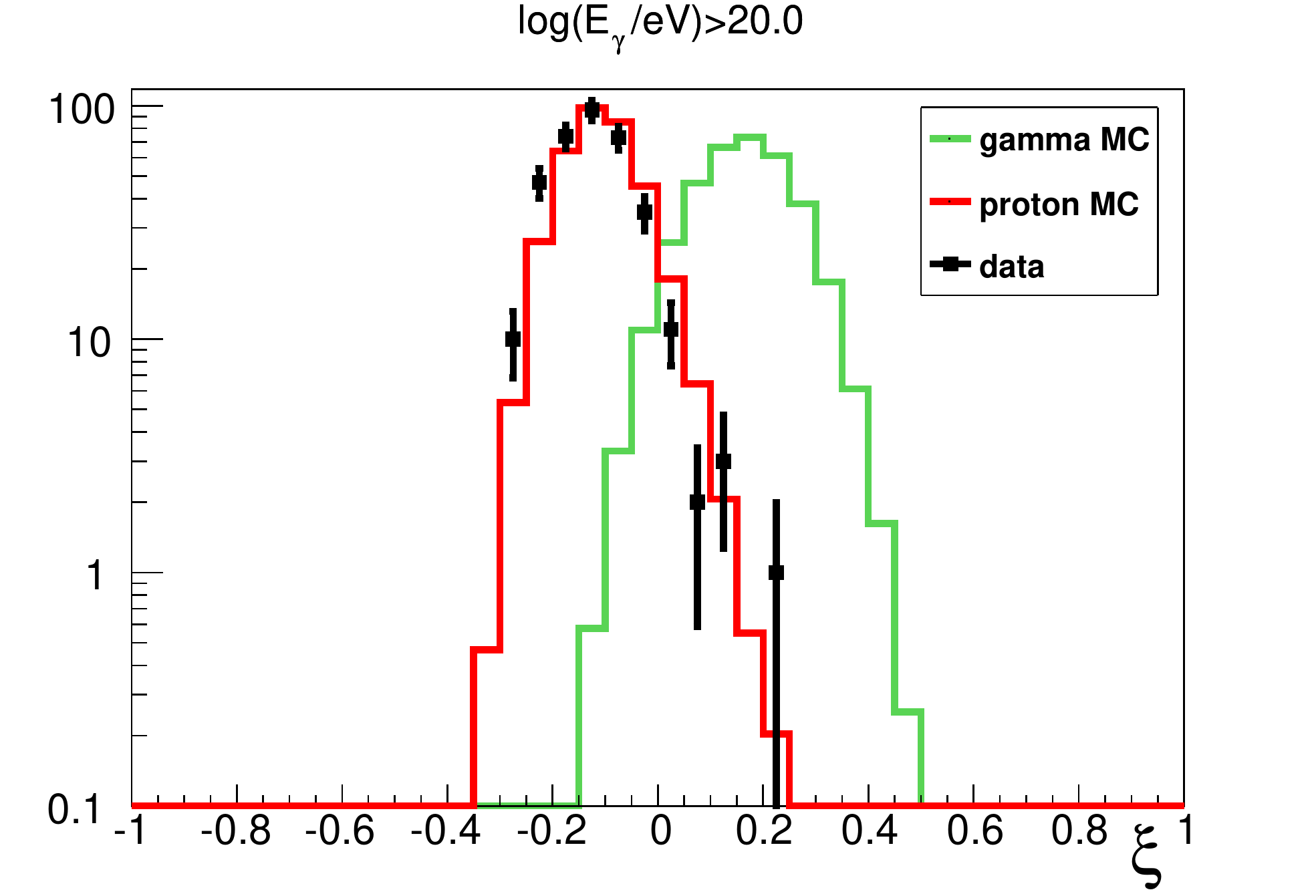}
\caption{\label{xi_distribution} The distributions of the photon and proton Monte-Carlo and data events
over the $\xi$ parameter for the five energy ranges (solid red --- protons, solid green --- photons, black dots --- data). }
\end{center}
\end{figure*}

The analysis method used in this study to distinguish between photon and proton
events is a boosted decision tree (BDT) classifier built with the 16 observable parameters
listed in the previous section. As an implementation of this method, we use the AdaBoost
algorithm~\citep{Freund1997119} from the TMVA package~\citep{Hocker:2007ht}
for ROOT~\citep{Brun:1997pa}, in the same way as in the recent TA studies~\citep{Abbasi:2018ywn, Abbasi:2018wlq}.

The BDT is trained to separate proton MC events from
photon MC events. Both proton and photon MC sets are split into
three parts with equal amount of events in each: 
one for training the classifier, the second one for testing the classifier and the last one for the calculation
of proton background and photon effective exposure, respectively. We train the classifier
separately in five photon energy ranges:
$E_\gamma>10^{18}$~eV, $E_\gamma>10^{18.5}$~eV, $E_\gamma>10^{19}$~eV, $E_\gamma>10^{19.5}$~eV and $E_\gamma>10^{20}$~eV.
As a result of the BDT procedure, the single multivariate analysis (MVA) parameter $\xi$ 
is assigned to each MC and data event. $\xi$ is defined to take values in the range $-1 < \xi < 1$,
where proton-induced events tend to have negative $\xi$ values,
and photon-induced events --- positive $\xi$ values. 
The resulting $\xi$ distributions of the MC events from the testing
sets and the data events for  all  considered energy ranges are shown in Fig.~\ref{xi_distribution}.

From Fig.~\ref{xi_distribution}, which shows the distributions of data and Monte-Carlo
irrespective of the direction in the sky, one can observe no deviation from
the proton distribution in the expected photon signal region. However,
possible excesses in one or several separate directions in the sky could
be overlooked if we analyze the all-sky averaged $\xi$ distribution. 
Hereafter we discuss method to set photon--flux upper limit and to search
for photon excesses from separate directions  on the sky  and present respective results.

It is important to note that at primary energies of order EeV and higher there is a potential systematic uncertainty in the
estimation of the hadron background for the photon signal. The bulk of the events are induced by protons and/or nuclei,
but their mass composition is not known precisely~\citep{Aab:2017cgk, Abbasi:2018nun, Abbasi:2018wlq}.
We have examined $\xi$ distributions of the iron nucleus-induced events and found that 
in average the iron-induced events are less "photon-like" than proton-induced events.
The results of TA work~\citep{Abbasi:2018wlq}, where the similar BDT-classifier was used, implies
that a mixed nuclei $\xi$ distribution would also deviate from the photon 
$\xi$ distribution stronger than the proton $\xi$ distribution.
Therefore,  the assumption  of the proton background for the photon
search is conservative.  However, we also perform independent photon search assuming more realistic
mixed nuclei background inferred form the TA SD data in our study~\citep{Abbasi:2018wlq}.

\subsection{Photon-flux upper limit }
\label{ul}

\begin{figure*}
\begin{center}
 \includegraphics[width=0.9\columnwidth]{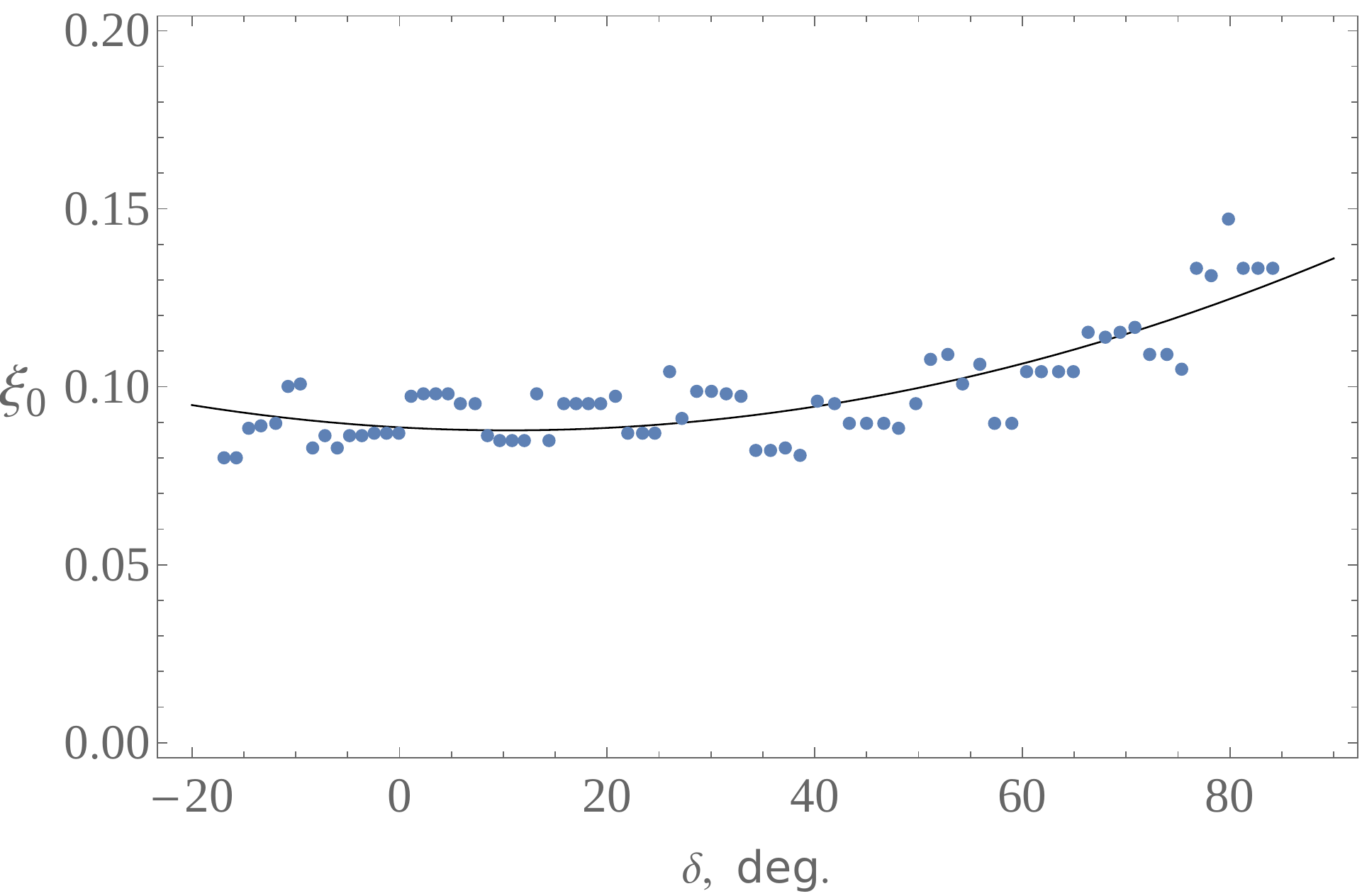}
 \includegraphics[width=0.9\columnwidth]{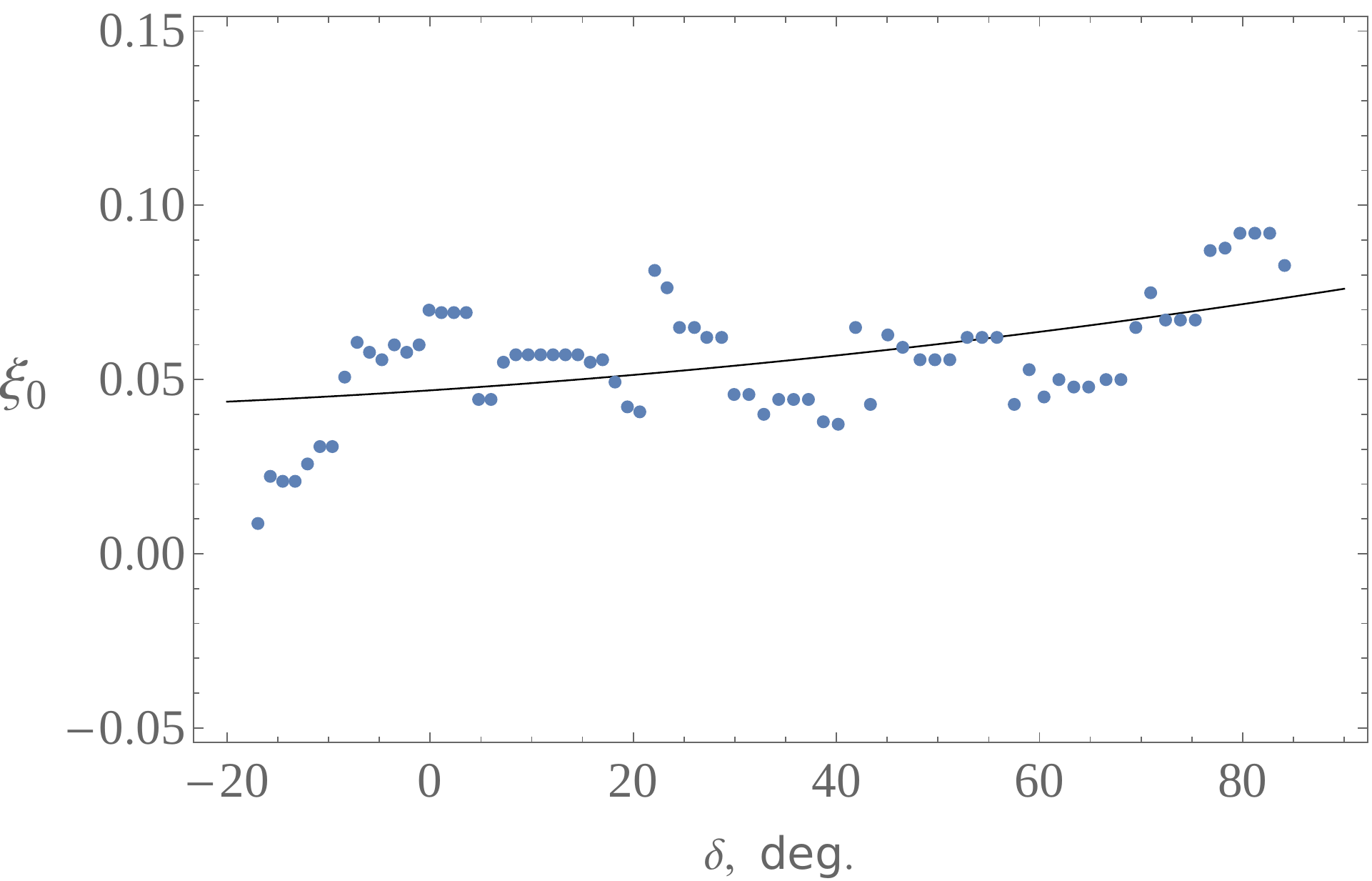}
\caption{
\label{xi_fit}
Examples of $\xi_0$ position as a function of declination and its smooth fitting 
for $E_\gamma > 10^{18}$~eV (left panel) and $E_\gamma > 10^{19}$~eV (right panel) photons.
Blue points are the cut positions obtained with the optimization of the photon flux-upper limit Eq.~(\ref{ul_formula})
in the respective declination bands. 
}
\end{center}
\end{figure*}

In general, the flux upper limit for the particular type of primaries is defined as:
\be
F_{\rm UL} = \frac{\mu_{\rm FC}(N_{\rm obs}, N_{\rm bg})}{A_{\rm eff}}
\label{ul_formula}
\ee
where $N_{\rm obs}$ is the number of detected events of a 
given type in a given energy range,
$N_{\rm bg}$ is the estimated number of background events in the same energy range,
$\mu_{\rm FC}$ is the upper bound of the respective Poisson mean for the given
confidence level, defined according to ~\citep{Feldman:1997qc},
and $A_{\rm eff}$ is the effective exposure of the experiment for the given type of primaries
in the same energy range.

In the present upper-limit calculation we assume the ``null hypothesis'',
i.e. that there is actually no photons and any excess counts from the expected background,
$N_{\rm obs} - N_{\rm bg}$, is considered as a fluctuation of background.

We consider two options of the background estimation. First one is $N_{\rm bg} = 0$, this assumption is conservative since
for the fixed $N_{\rm obs}$ the upper-limit value is higher for a lower value of $N_{\rm bg}$.
Second one is a ``real'' background of mixed nuclei with the mean $\ln A$ following
the one derived from the same TA SD data with the same MVA method in our work~\citep{Abbasi:2018wlq}.
This background is estimated by down-scaling of the proton background to the respective mean $\ln A$,
linearly with $\ln A$, taking into account the recaling of SD energy scale
used in work~\citep{Abbasi:2018wlq} to the $E_\gamma$ energy scale used in this work.

The separation between photon and proton primaries is defined by a 
cut on MVA-variable $\xi$. The cut is set at some value $\xi_0$
so that any proton with $\xi > \xi_0$ is considering as a photon candidate and any
photon with $\xi > \xi_0$ is contributing to the effective exposure.

To find the minimum value of $F_{\rm UL}^\gamma$ as a function of $\xi_0$ 
we optimize the cut position assuming:
$N_{\rm obs} = N_{\rm p}(\xi > \xi_0)$,

where $N_{\rm p}(\xi > \xi_0)$ is the number of protons passing the $\xi$-cut.
As one can see from Fig.~\ref{xi_distribution}, the number of MC photon events
passing the $\xi$-cut is decreasing with the growth of $\xi_0$ leading to the respective
decrease of the exposure $A_{\rm eff}^\gamma$, also the number of photon candidates,
$N_{\rm obs} = N_{\rm p}(\xi > \xi_0)$, is decreasing, 
but $N_{\rm obs}=0$ yields a constant non-zero value of $\mu_{\rm FC}$.
This implies that there indeed should be a non-trivial minimum value of $F_{\rm UL}^\gamma$
as a function of $\xi_0$.

For the $\xi$-cut optimization we use the proton Monte-Carlo set normalized to the size of the data set
and the photon Monte-Carlo set, the latter is used for the calculation of the photon effective
exposure $A_{\rm eff}^\gamma$. It is important to note that the optimization procedure tends
to place $\xi_0$ at the right edge of the proton distribution in Fig.~\ref{xi_distribution}, therefore
the number of candidates $N_{\rm obs}$ and the upper-limit value $F_{\rm UL}$ are subject to fluctuations.
These fluctuations become apparent when one considers upper limits
for particular directions in the sky with small number of events.

Up to this moment the procedures of upper-limit calculation and cut optimization were 
similar to those used to search for diffuse photons~\citep{Abbasi:2018ywn}.
The difference of the analysis procedure used here from that of~\citep{Abbasi:2018ywn}
is in the usage of the separate event sets for different directions in the sky.
The $\xi$-cut is also optimized separately for every direction studied.
We pixelize the sky in equatorial coordinates $\{\alpha, \delta\}$ using the HEALPix package~\citep{Gorski:2004by} into 12288
pixels ($N_{side}=32$). For the pixel ``i'' with the center $\{\alpha_i, \delta_i\}$
the corresponding data set contains events located inside a spherical cap region around
the pixel center within an angular distance that equals to the experiment's 
angular resolution at the respective energy
(see Tab.~\ref{gamma_angles})$^{1)}$\let\thefootnote\relax\footnote{$^{1)}$ The distance between any pixel centers
is smaller that experiment's angular resolution at all considered energies, therefore the experiment FOV is overlapped without gaps,
but some events in adjacent pixels could be the same.}.

The effective exposure of the experiment to photons at the pixel ``i'' is given by:
\be
A_{\rm eff}^i = S \cdot T \cdot \overline{\cos{\theta_i}} \frac{N^i_{MC, \gamma}(\xi > \xi_0)}{N^i_{MC, \gamma}}
\label{p_exposure}
\ee
where $S$ is the area of the experiment, $T$ is the period of observation, $\theta_i$ is
the zenith angle at which the pixel ``i'' is seen by the experiment, $N^i_{MC, \gamma}$ is the
total number of photon events simulated in the respective pixel and $N^i_{MC, \gamma}(\xi > \xi_0)$
is the number of these events that pass the $\xi$-cut. The same pixel in equatorial coordinates
is seen by the experiment at different $\theta$ depending on time, therefore the
diurnal mean value $\overline{cos{\theta}}$ is used. It is given by the expression~\citep{Sommers:2000us}:
\be
\overline{cos{\theta}} = \cos \lambda_0\,\cos\delta\,\sin\alpha_m+\alpha_m\sin \lambda_0\,\sin\delta,
\label{sommers_omega}
\ee
where $\delta$ is the declination, $\lambda_0$ is the geographical latitude
of the experiment, $\theta_{max}$ is the maximum
zenith angle of the events considered in the particular analysis and $\alpha_m$ is given by the expression
\be
\alpha_m=\begin{cases}
0 & ;\zeta>1,\\
\pi & ;\zeta<-1,\\
\arccos\zeta & ; -1 < \zeta < 1\,;
\end{cases}
\ee
where
\be
\zeta = \frac{(\cos\theta_\text{max}-\sin \lambda_0\,\sin\delta)}{\cos \lambda_0\,\cos\delta}.
\ee

The ``effective'' part of the exposure, $\frac{N^i_{MC, \gamma}(\xi > \xi_0)}{N^i_{MC, \gamma}}$,
is calculated using the photon Monte-Carlo. 
To have enough statistics for this calculation, 
one needs to generate separate Monte-Carlo sets for each sky-map pixel. However,
it is technically unreasonable, since the exposure depends only on declination of the given pixel.
We use the following method to increase the Monte-Carlo statistics in each pixel:
$\xi_0$ is optimized over the events belonging to the whole constant--declination
band whose width is twice the angular resolution centered in the given pixel. 
This method resembles the so-called ``scrambling technique''~\citep{Cassiday:1989yw}, which was used for instance
in the Pierre Auger Observatory search for photon point sources~\citep{Aab:2014bha}. The additional advantage
of the used method is the preservation of relatively large effective statistics of the
Monte-Carlo events in each pixel, including the variety over $\xi$--parameter. 
We have found that in this case, fluctuations of the $\xi_0$ position
between adjacent pixels are smaller, compared to the standard scrambling technique.
It is reasonable to smooth these fluctuations even further by making a least-squares 
fit of a $\xi_0$ position as a function of declination with a smooth function, for which
we use a second--order polynomial. 
As it has been mentioned before, the flux upper limit remains conservative after this operation.
The examples of $\xi_0$ as a function of declination and its smooth fitting are shown
in Fig.~\ref{xi_fit}$^{2)}$\let\thefootnote\relax\footnote{$^{2)}$ The
$\xi_0$ points for adjacent declination bands are clustered because these
bands are overlapping with each other and a part of their MC events is one and the same.}.

As the $\xi_0$ position for the pixel ``i'' is fixed, the actual upper-limit value is
calculated using the definition~(\ref{ul_formula})  with $N^i_{\rm obs} = N^i_{\rm data}(\xi > \xi_0)$,
where $N^i_{\rm data}$ is a number of data events belonging to the respective pixel and $N^i_{\rm bg} = 0$
or $N^i_{\rm bg} = N^i_{\ln A}(\xi > \xi_0)$, where $N^i_{\ln A}$ is a number of ``real'' background events. 

The Telescope Array field of view for the considered zenith angle cut ($0^\circ < \theta < 60^\circ$) spans from
$-20.7^\circ$ to $90^\circ$ in declination. However, the event statistics is low in the constant--declination 
bands near the edges of this interval. Therefore we reduce the considered sky region to $-15.7^\circ \leq \delta \leq 85^\circ$.
It contains  7848 pixels. 

\subsection{Results}
\label{res}

\begin{figure*}
\begin{center}
 \includegraphics[width=0.9\columnwidth]{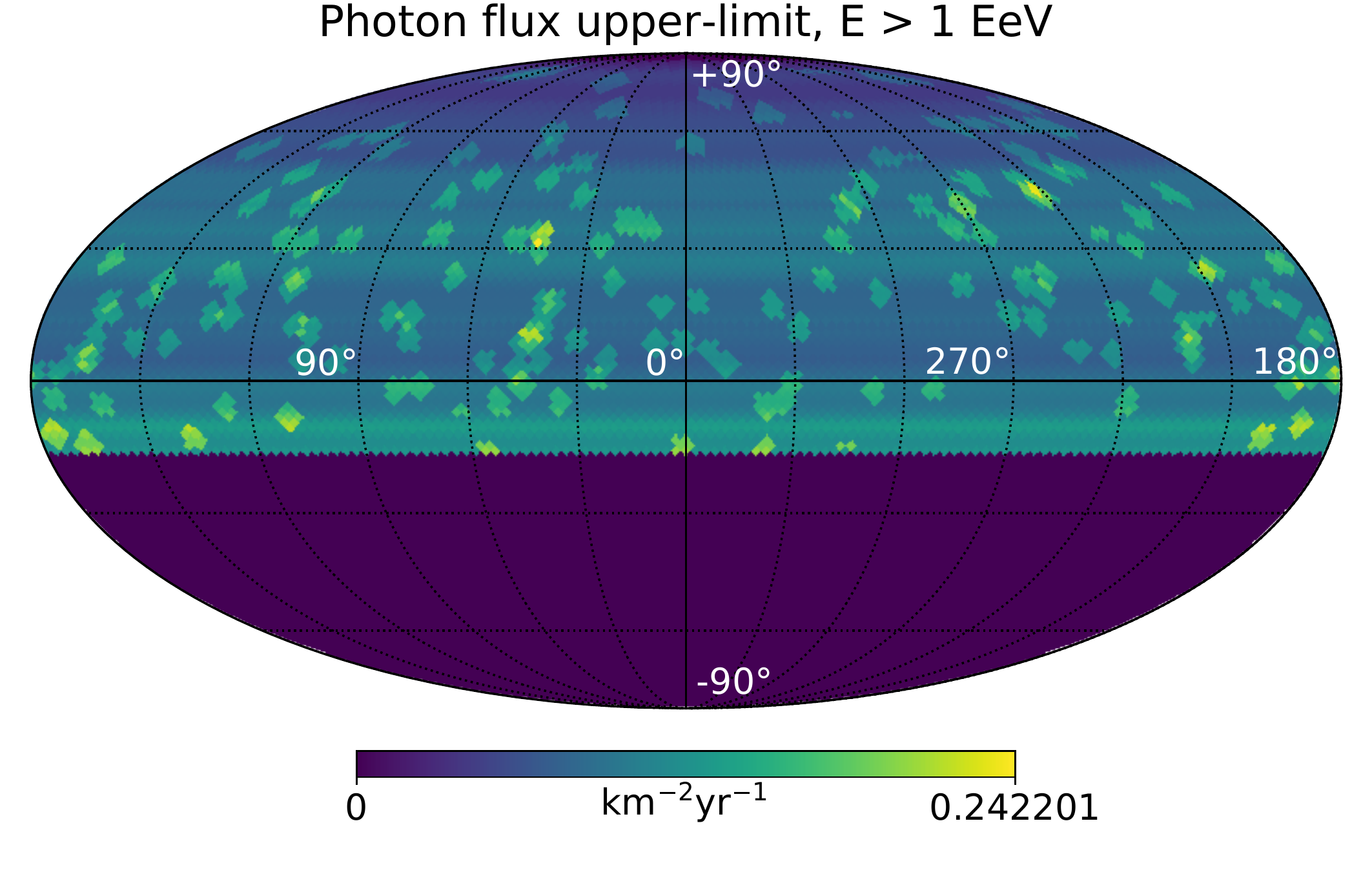}
 \includegraphics[width=0.9\columnwidth]{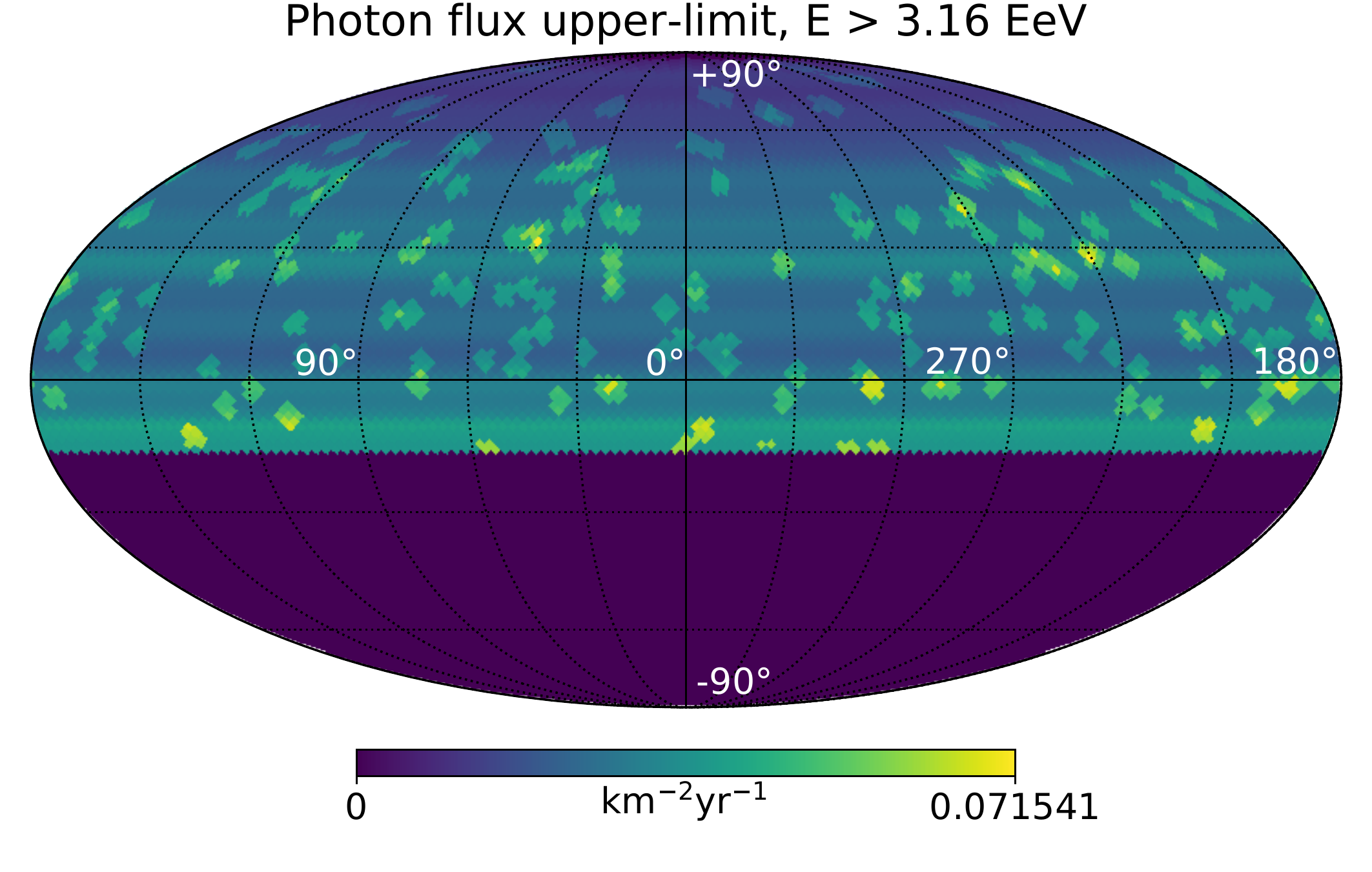}\\
 \includegraphics[width=0.9\columnwidth]{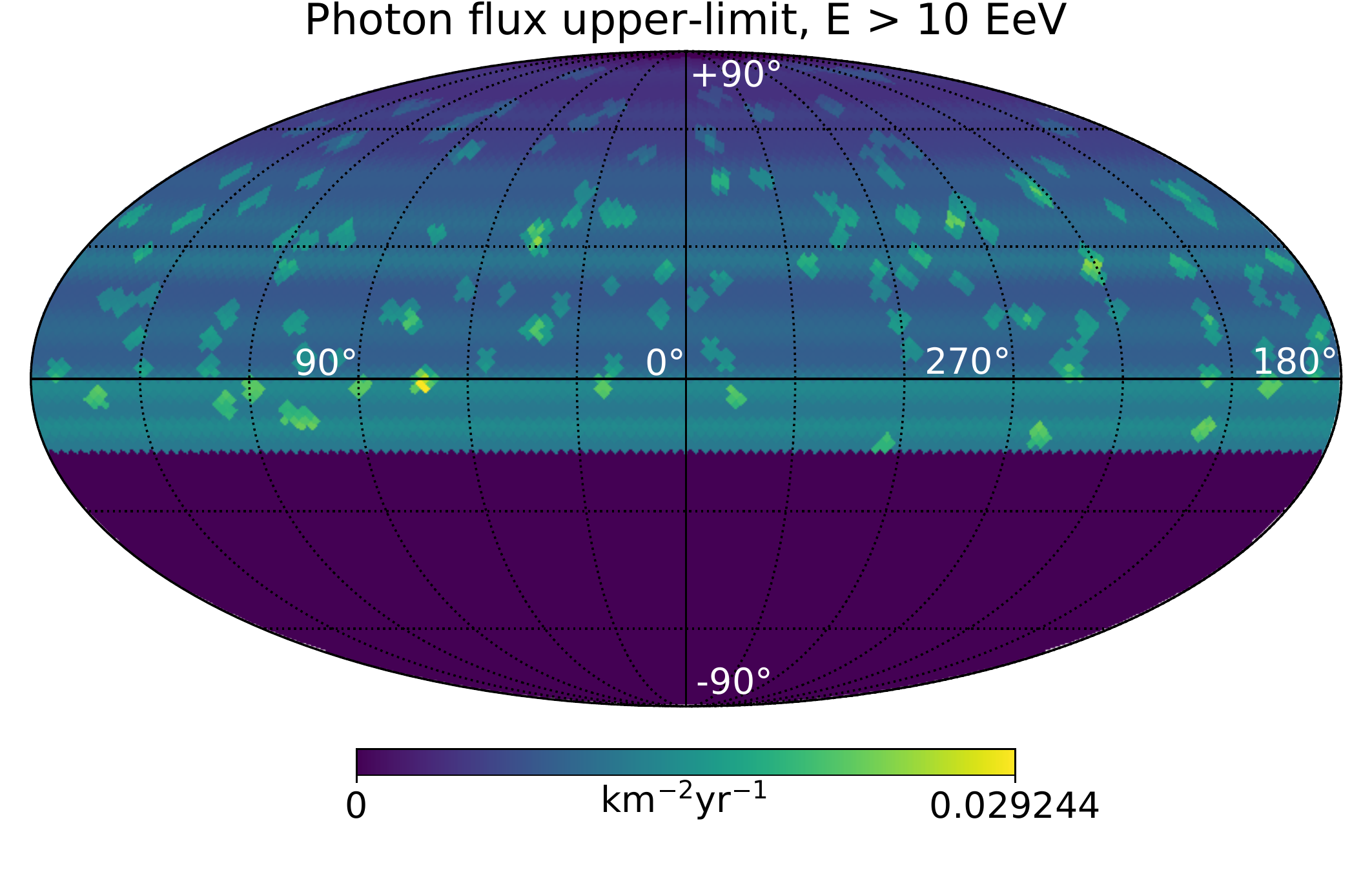}
 \includegraphics[width=0.9\columnwidth]{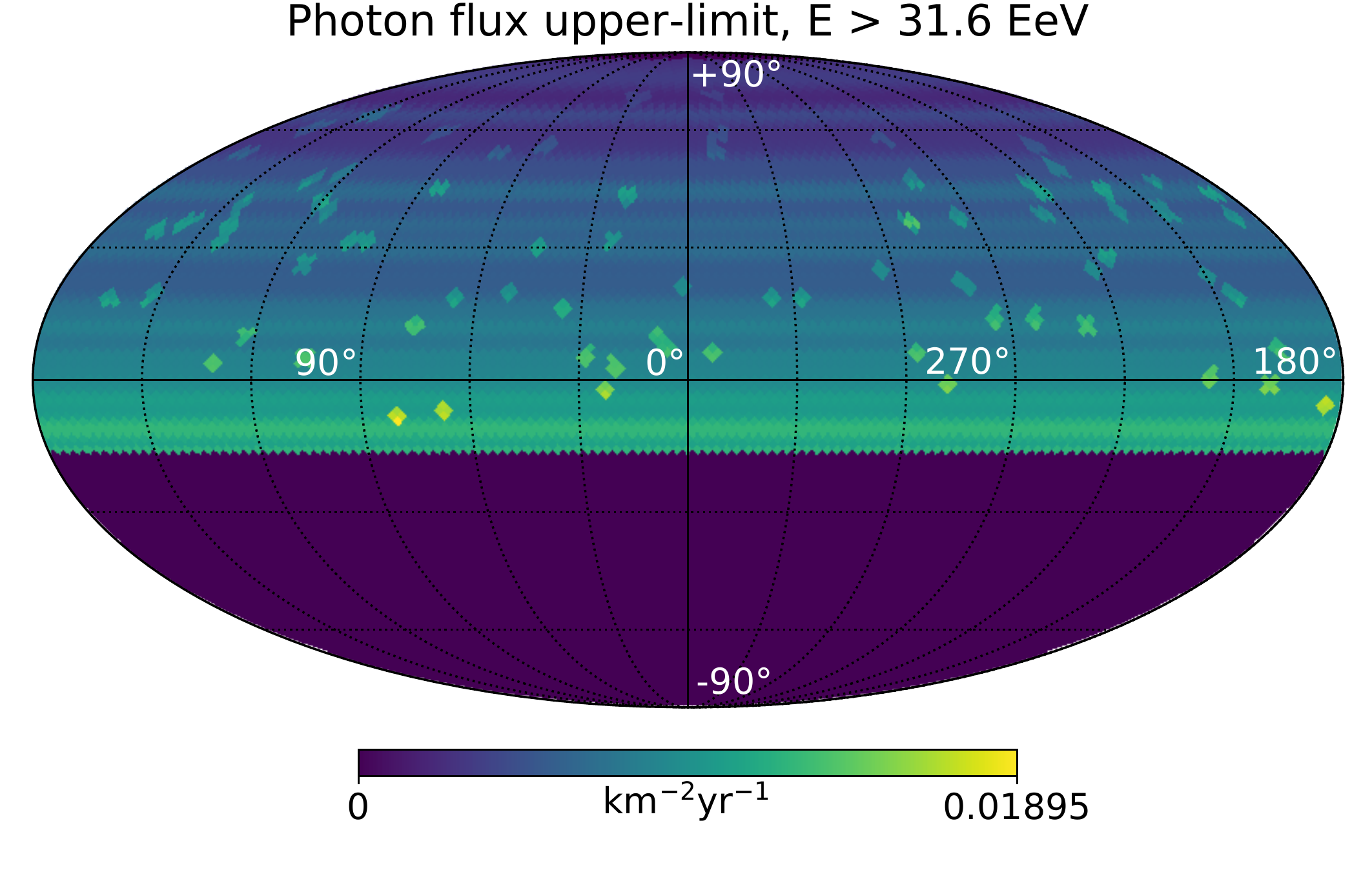}\\
 \includegraphics[width=0.9\columnwidth]{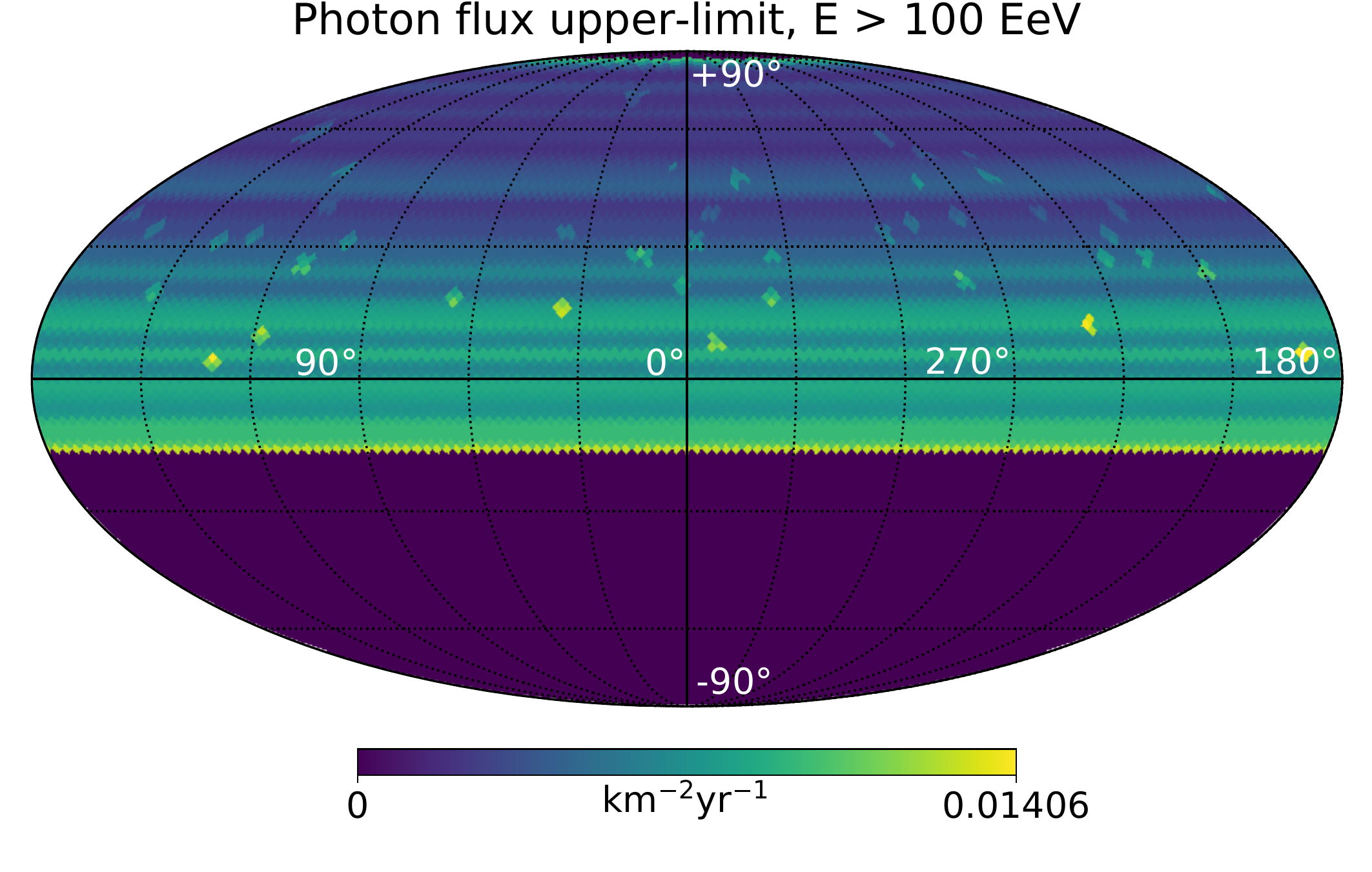}
\caption{
\label{limits}
Maps of point-source photon flux upper limits (95\% C.L.) for various photon energies  calculated in
zero background assumption and  plotted in equatorial coordinates.
}
\end{center}
\end{figure*}

\begin{table*}
\begin{center}

\begin{tabular}{|c|c|c|c|c|}
\hline
$E_\gamma$, eV & $\langle F_\gamma \rangle \leq$, ${\rm km}^{-2} {\rm yr}^{-1}$ (zero bg.) & $\langle F_\gamma \rangle \leq$, ${\rm km}^{-2} {\rm yr}^{-1}$ (``real'' bg.) & $\langle N_{\rm bg} \rangle$ & max. $\gamma$ signif. (pre-trial) \\
\hline
$> 10^{18.0}$ & 0.094 & 0.069 & 0.49 & $2.72\; \sigma$ \\ \hline
$> 10^{18.5}$ & 0.029 & 0.021 & 0.52 & $2.71\; \sigma$ \\ \hline
$> 10^{19.0}$ & 0.010 & 0.0074 & 0.34 & $2.89\; \sigma$ \\ \hline
$> 10^{19.5}$ & 0.0071 & 0.0055 & 0.10 & $2.76\; \sigma$ \\ \hline
$> 10^{20.0}$ & 0.0058 & 0.0045 & 0.029 & $3.43\; \sigma$ \\ \hline
\end{tabular}

\end{center}
\caption{
Point-source photon-flux upper limits and proton backgrounds averaged over all pixels
together with the maximum pre-trial significance of the photon excess over proton background. 
}
\label{avg_lim}
\end{table*} 

The  95\% C.L. photon-flux upper limits calculated in
zero background assumption  for each pixel in the Telescope Array field of view and for
various photon energies are shown in Fig.~\ref{limits}. The numerical values of these limits
 as well as the limits calculated with the ``real'' background assumption 
are  given  in the supplementary material. The values of the limits averaged
over all pixels are presented in Table~\ref{avg_lim}.

\begin{figure*}
\begin{center}
 \includegraphics[width=0.9\columnwidth]{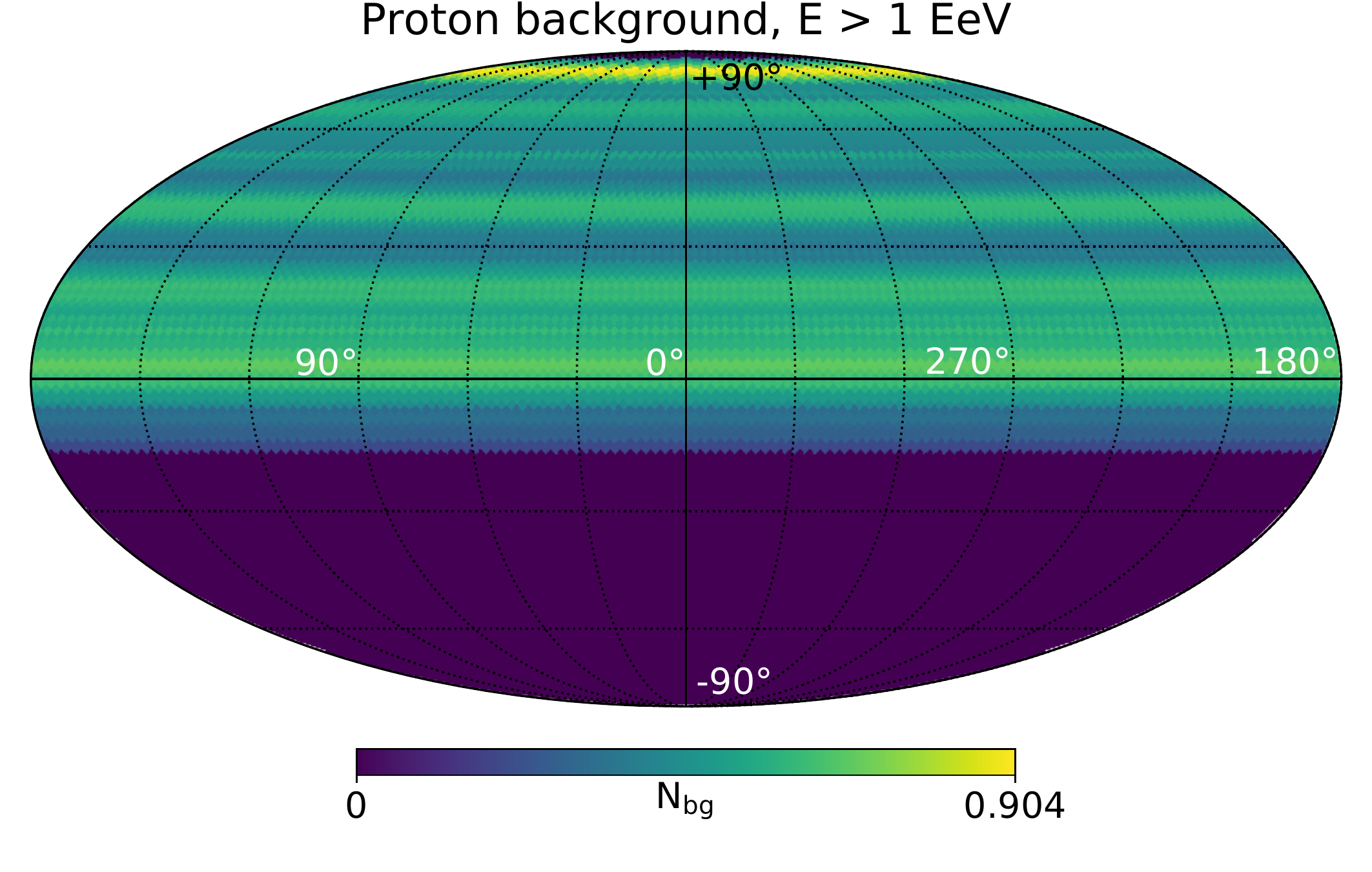}
 \includegraphics[width=0.9\columnwidth]{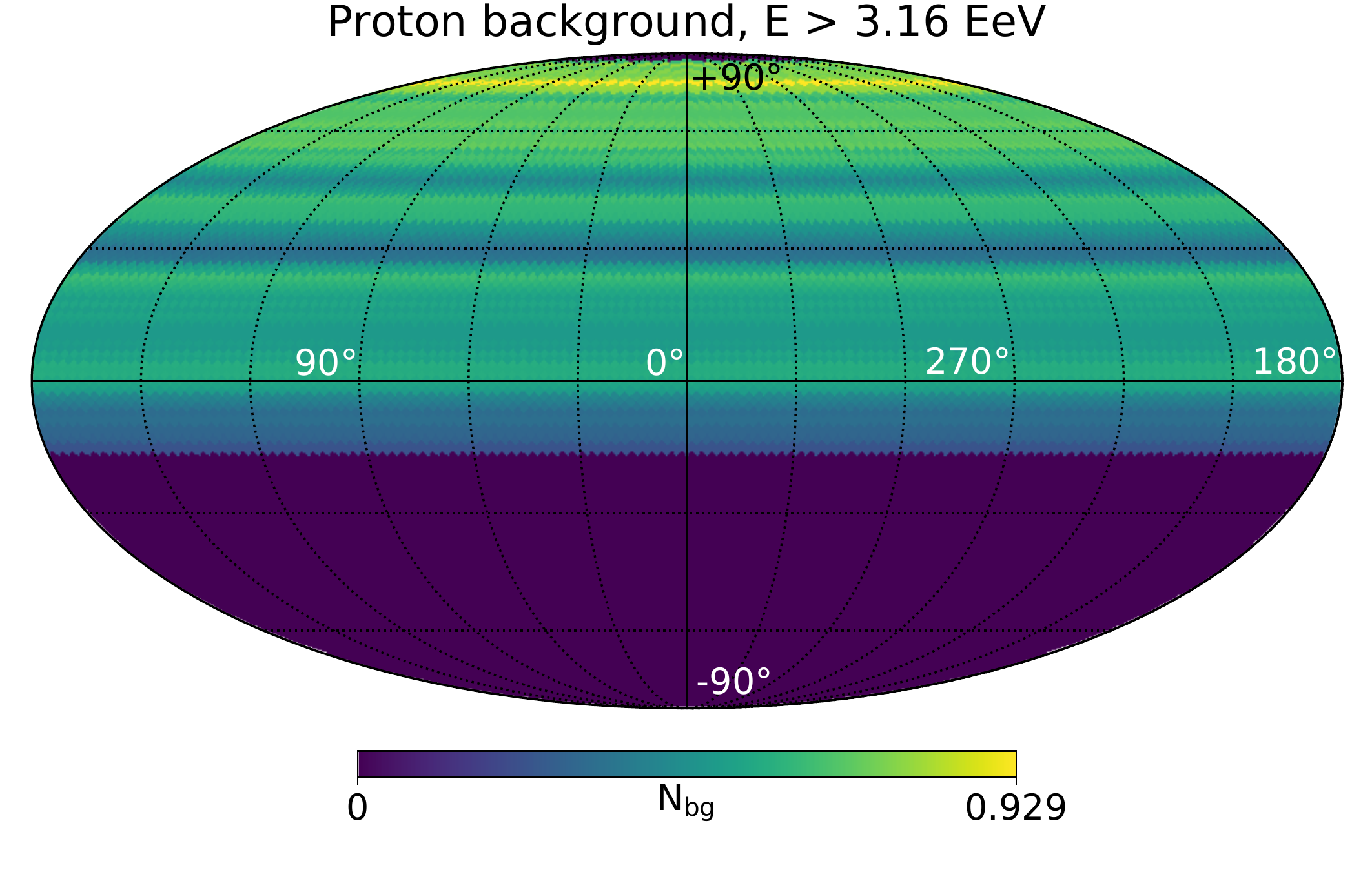}\\
 \includegraphics[width=0.9\columnwidth]{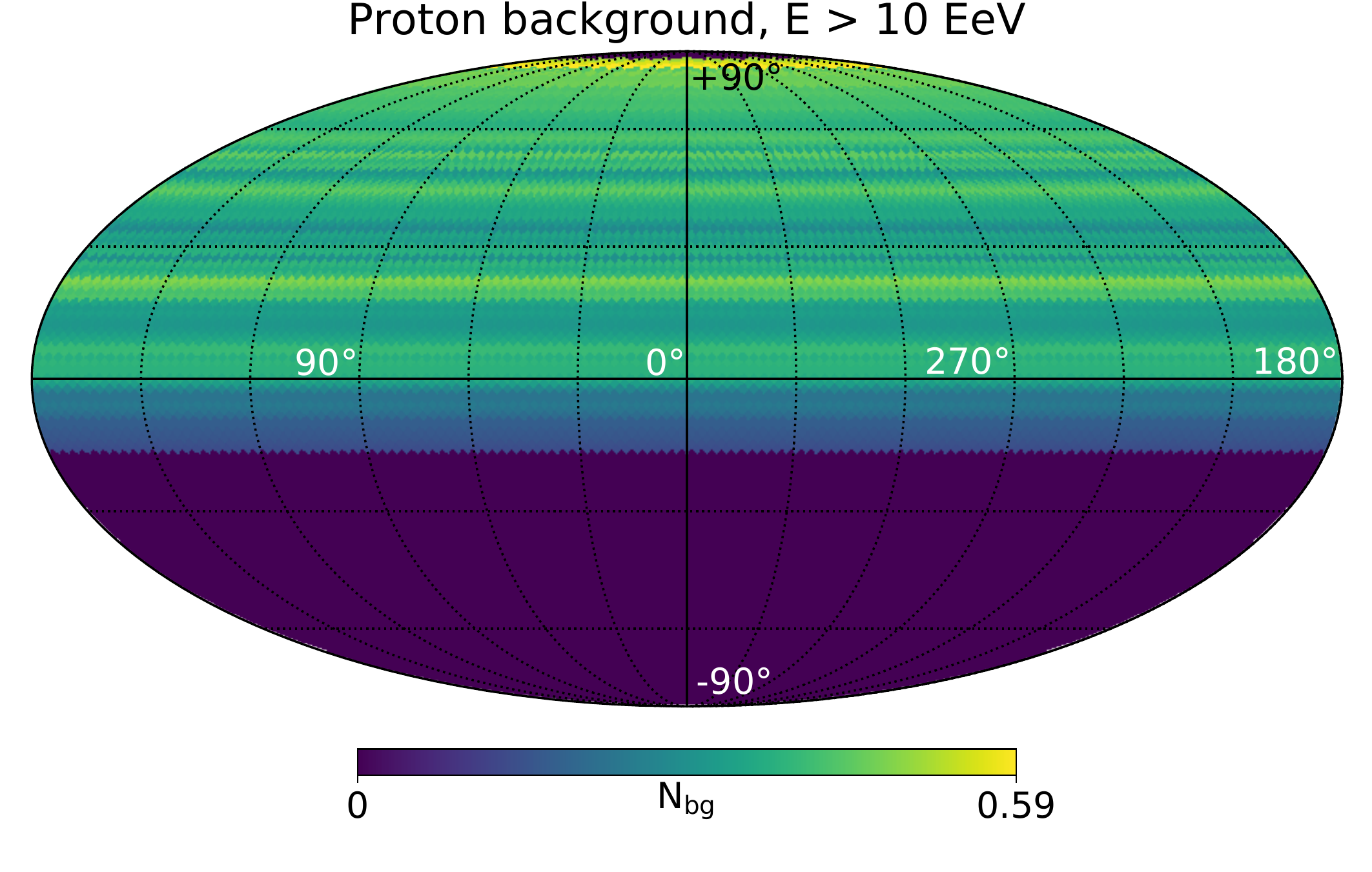}
 \includegraphics[width=0.9\columnwidth]{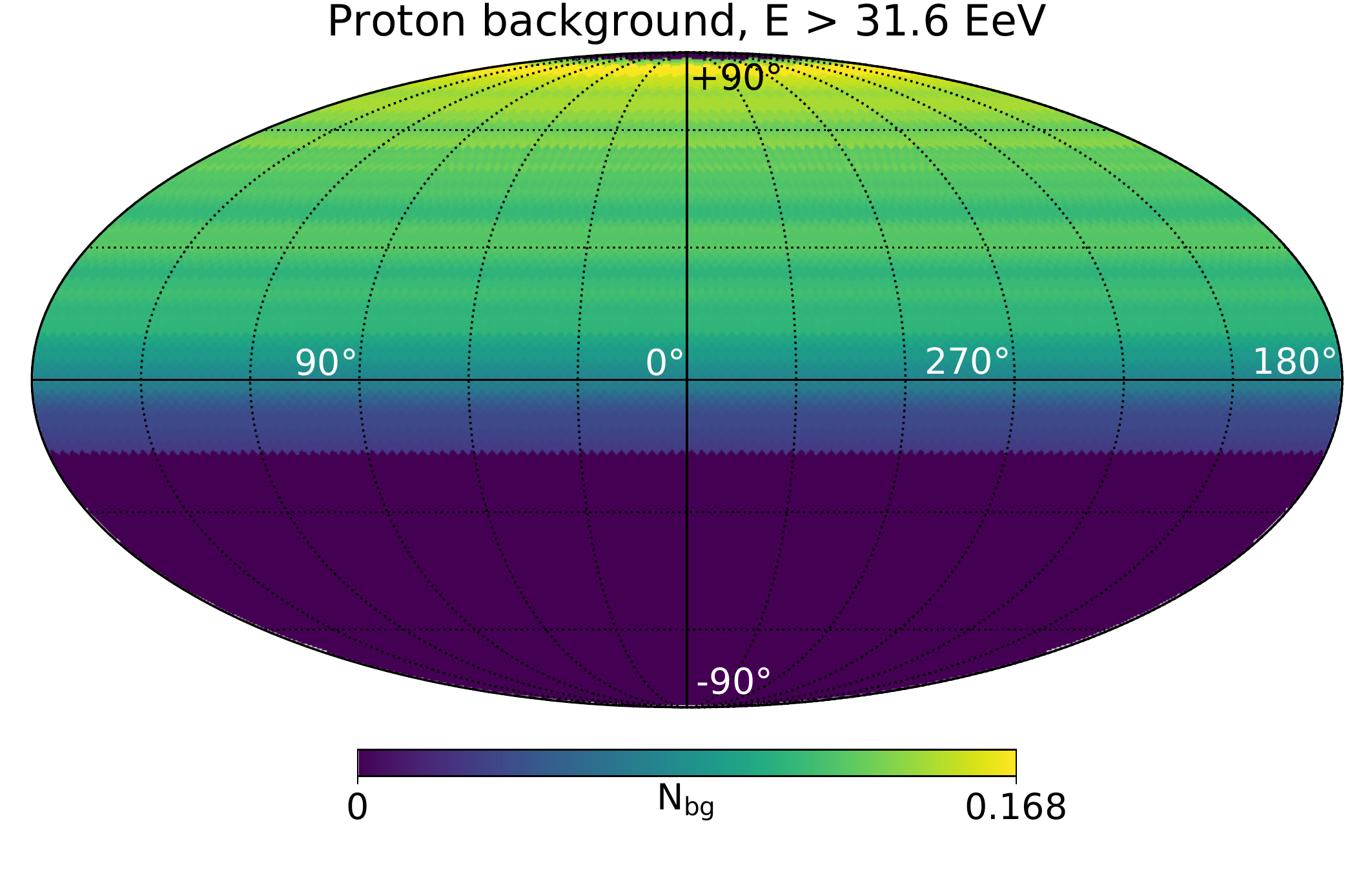}\\
 \includegraphics[width=0.9\columnwidth]{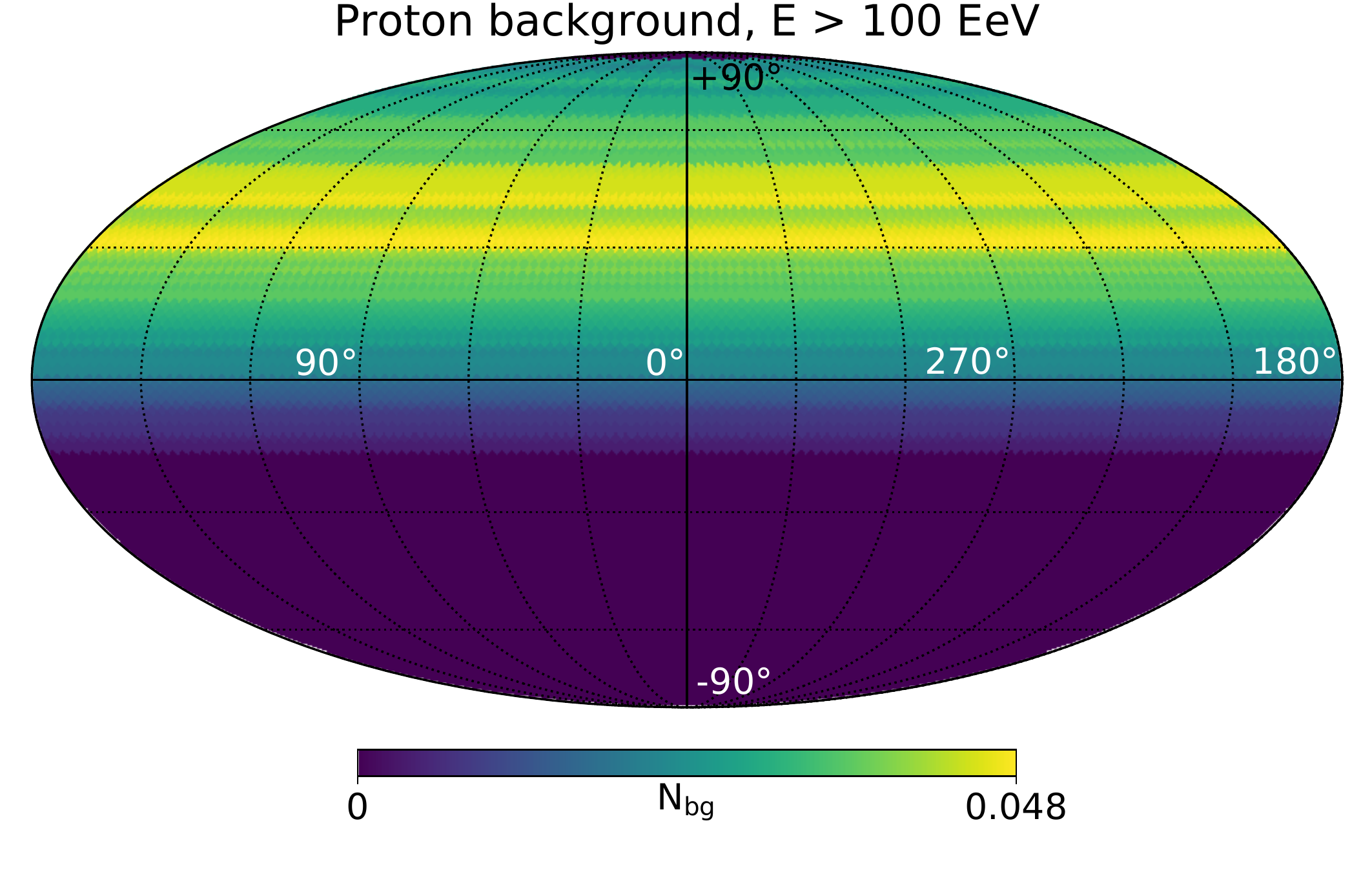}
\caption{\label{background} The distributions of the numbers of proton background events over the sky map for the various photon energies
plotted in equatorial coordinates.}
\end{center}
\end{figure*}

The null hypothesis assumed for the photon upper-limit calculation is
not optimal for the photon search. However
the rough estimation of the possible photon signal could be made already in this setup.
We optimize $\xi_0$ in each declination band with the same assumptions as in the previous section, 
and estimate the background in each pixel as the appropriately normalized number of protons
that pass the cut: $N^i_{MC, p}(\xi > \xi_0)$. 
For the photon excess calculation, the assumption of proton background is conservative
as it should be higher than any mixed nuclei background, as it was discussed in Sec.~\ref{mva}. 
The background maps for various photon energies are shown in Fig.~\ref{background}.
 The maxima among all pixels pre-trial photon candidate
excesses over the proton background are presented in the Table~\ref{avg_lim} along with the
average values of the proton background. 
The highest pre-trial excess significance, $3.43 \sigma$ ($N_{\rm bg}=0.036$ and $N_{\rm obs} = 2$), 
appears in the highest energy bin $E_\gamma > 10^{20}$~eV, at $\{\alpha = 155.3^\circ, \delta = 60.4^\circ\}$ pixel. 
To make a simple estimation of the post-trial p-value one can
use the Bonferroni correction, i.e. to multiply the number of trials by the mimimum pre-trial excess p-value~\citep{Miller:1991}.
In turn, the number of trials could be estimated as the number of non-overlapping pixel-size
regions of the map which is several times smaller than the actual number of pixels.
The resulting post-trial significances estimated 
in this way appear to be below $1 \sigma$ level for all points of the sky at all considered energies.
Therefore we conclude that, at the present level of point-source photon search
sensitivity, there is no evidence for the photon signal. The actual results for
each sky-map pixel at various energies are given in the text files supplemented
to this paper. The format of the files is described in the Supplementary section.

The main systematic uncertainties for the photon-flux upper limits
are related to the overestimation of the $E_\gamma$ parameter for hadron-induced events
and to the uncertainty of the primary hadron mass-composition.
The former uncertainty leads to the overestimation of the hadron background
and subsequently to the looser photon-flux upper limit.  As for the hadron mass composition
uncertainty, the assumption of the proton composition which we used for the $\xi$-cut optimization
could only make the photon-flux upper limit looser comparing to a mixed nuclei composition case.
Therefore, the limits set are conservative with respect to the both of these uncertainties.

Finally, the last assumption that affects the result is the assumption of the background in (Eq.~\ref{ul_formula}).
The most conservative limits are set for zero background assumption, while the ``real''
mixed nuclei background assumption yields somehow more realistic limits.

\section{Discussion and Conclusions}
\label{dis}
The upper limits are set on the fluxes of photons from each particular direction in the sky 
in the TA field of view, according to the experiment's angular resolution with respect to photons. 
The only results of the ultra-high energy point-source photon flux  upper limits were
presented so far by the Pierre Auger experiment~\citep{Aab:2014bha, Aab:2016bpi}.  The
comparison of those results to ours is not  straightforward  as the photon energy
range of the Auger search, $10^{17.3}<E_\gamma<10^{18.5}$~eV, does not fully coincide 
to any of our ranges of search.
Regardless of that,  the average point-source photon flux upper-limit of Auger
$\langle F_{\gamma} \rangle \leq 0.035\; {\rm km}^{-2} {\rm yr}^{-1}\;,$ is from two to three times
lower than our average limit  for the $E_\gamma > 10^{18}$~eV.
The results for the energies larger than $E > 10^{18.5}$~eV are obtained here for the first time.

The point-source photon-flux upper limits derived in the present study
can be used to constrain various models of astrophysics and particle physics.
One can assume a distribution of photon sources and impose the constraints on
their properties using the combination of point-source limits. In principle these
constraints could be stronger than those derived  from  the diffuse photon-flux limits.
The models that could be probed with the present photon point-source flux
limits include cosmogenic photon generation models as well as top-down models of ultra-high energy
photons production such as heavy decaying dark matter. 

\section*{Supplementary}
Photon point-source flux upper limits and photon excess pre-trial
significances for all sky-map pixels are summarized in the separate
file for each energy bin, named ``limit\_$[\log(E_\gamma/{\rm eV})]$.txt''. 
The file contains several columns with the following data. 

Column 1: HEALpix pixel number (RING, started from 24)

Column 2: pixel $\alpha$, rad. 
 
Column 3: pixel $\delta$, rad. 

Column 4: $\xi$-cut value

Column 5: proton background value

Column 6: number of $\gamma$-candidate events

Column 7:  95\% C.L. $F_\gamma$ upper limit, ${\rm km}^{-2} {\rm yr}^{-1}$ (zero background assumption)

Column 8: 95\% C.L. $F_\gamma$ upper limit, ${\rm km}^{-2} {\rm yr}^{-1}$ (``real'' background assumption) 

Column 9: pre-trial $\gamma$ excess p-value 

Column 10: pre-trial $\gamma$ excess significance

As it was mentioned in Sec.~\ref{res} the proton background value is used only for the calculation of
photon excess p-value and significance, while the  upper limits are calculated in either zero background assumption
or ``real'' mixed nuclei background assumption.  For pixels with the number of $\gamma$-candidates
less than $p$ background both p-value and significance are set to zero.

\section*{Acknowledgments}
The Telescope Array experiment is supported by the Japan Society for
the Promotion of Science (JSPS) through 
Grants-in-Aid
for Priority Area
431,
for Specially Promoted Research 
JP21000002, 
for Scientific Research (S) 
JP19104006, 
for Specially Promoted Research 
JP15H05693, 
for Scientific Research (S)
JP15H05741, for Science Research (A) JP18H03705 and 
for Young Scientists (A)
JPH26707011; 
by the joint research program of the Institute for Cosmic Ray Research (ICRR), The University of Tokyo; 
by the U.S. National Science
Foundation awards PHY-0601915,
PHY-1404495, PHY-1404502, and PHY-1607727; 
by the National Research Foundation of Korea
(2016R1A2B4014967, 2016R1A5A1013277, 2017K1A4A3015188, 2017R1A2A1A05071429)
and Belgian Science Policy under IUAP VII/37 (ULB).
The development and application of the multivariate
analysis method is supported by the Russian Science Foundation grant
No. 17-72-20291 (INR).
The foundations of Dr. Ezekiel R. and Edna Wattis Dumke, Willard L. Eccles, and George
S. and Dolores Dor\'e Eccles all helped with generous donations. The
State of Utah supported the project through its Economic Development
Board, and the University of Utah through the Office of the Vice
President for Research. The experimental site became available through
the cooperation of the Utah School and Institutional Trust Lands
Administration (SITLA), U.S. Bureau of Land Management (BLM), and the
U.S. Air Force. We appreciate the assistance of the State of Utah and
Fillmore offices of the BLM in crafting the Plan of Development for
the site. Patrick Shea assisted the collaboration with valuable advice 
on a variety of topics. The people and the officials of Millard County, 
Utah have been a source of
steadfast and warm support for our work which we greatly appreciate. 
We are indebted to the Millard County Road Department for their efforts 
to maintain and clear the roads which get us to our sites. 
We gratefully acknowledge the contribution from the technical staffs of
our home institutions. An allocation of computer time from the Center
for High Performance Computing at the University of Utah is gratefully
acknowledged.
The cluster of the Theoretical Division of INR RAS was
used for the numerical part of the work. The lightning data used in this paper was obtained
from Vaisala, Inc. We appreciate Vaisala's academic research policy.


\bibliographystyle{mnras}
\bibliography{ref}

\begin{thebibliography}{}
\makeatletter
\relax
\def\mn@urlcharsother{\let\do\@makeother \do\$\do\&\do\#\do\^\do\_\do\%\do\~}
\def\mn@doi{\begingroup\mn@urlcharsother \@ifnextchar [ {\mn@doi@}
  {\mn@doi@[]}}
\def\mn@doi@[#1]#2{\def\@tempa{#1}\ifx\@tempa\@empty \href
  {http://dx.doi.org/#2} {doi:#2}\else \href {http://dx.doi.org/#2} {#1}\fi
  \endgroup}
\def\mn@eprint#1#2{\mn@eprint@#1:#2::\@nil}
\def\mn@eprint@arXiv#1{\href {http://arxiv.org/abs/#1} {{\tt arXiv:#1}}}
\def\mn@eprint@dblp#1{\href {http://dblp.uni-trier.de/rec/bibtex/#1.xml}
  {dblp:#1}}
\def\mn@eprint@#1:#2:#3:#4\@nil{\def\@tempa {#1}\def\@tempb {#2}\def\@tempc
  {#3}\ifx \@tempc \@empty \let \@tempc \@tempb \let \@tempb \@tempa \fi \ifx
  \@tempb \@empty \def\@tempb {arXiv}\fi \@ifundefined
  {mn@eprint@\@tempb}{\@tempb:\@tempc}{\expandafter \expandafter \csname
  mn@eprint@\@tempb\endcsname \expandafter{\@tempc}}}

\bibitem[\protect\citeauthoryear{Aab et~al.}{Aab et~al.}{2014}]{Aab:2014bha}
Aab A.,  et~al., 2014, \mn@doi [Astrophys. J.] {10.1088/0004-637X/789/2/160},
  789, 160

\bibitem[\protect\citeauthoryear{Aab et~al.}{Aab et~al.}{2017a}]{Aab:2016bpi}
Aab A.,  et~al., 2017a, \mn@doi [Astrophys. J.] {10.3847/2041-8213/aa61a5},
  837, L25

\bibitem[\protect\citeauthoryear{Aab et~al.}{Aab et~al.}{2017b}]{Aab:2017cgk}
Aab A.,  et~al., 2017b, \mn@doi [Phys. Rev.] {10.1103/PhysRevD.96.122003}, D96,
  122003

\bibitem[\protect\citeauthoryear{Aab et~al.}{Aab et~al.}{2017c}]{Aab:2016agp}
Aab A.,  et~al., 2017c, \mn@doi [JCAP] {10.1088/1475-7516/2017/04/009}, 1704,
  009

\bibitem[\protect\citeauthoryear{Abbasi et~al.}{Abbasi
  et~al.}{2006}]{Abbasi:2005qy}
Abbasi R.~U.,  et~al., 2006, \mn@doi [Astrophys. J.] {10.1086/498142}, 636, 680

\bibitem[\protect\citeauthoryear{Abbasi et~al.}{Abbasi
  et~al.}{2015}]{Abbasi:2014wza}
Abbasi R.~U.,  et~al., 2015, \mn@doi [Astrophys. J.]
  {10.1088/0004-637X/804/2/133}, 804, 133

\bibitem[\protect\citeauthoryear{Abbasi et~al.}{Abbasi
  et~al.}{2017}]{ABBASI20172565}
Abbasi R.,  et~al., 2017, \mn@doi [Physics Letters A]
  {https://doi.org/10.1016/j.physleta.2017.06.022}, 381, 2565

\bibitem[\protect\citeauthoryear{Abbasi et~al.}{Abbasi
  et~al.}{2018a}]{Abbasi:2017muv}
Abbasi R.,  et~al., 2018a, \mn@doi [J. Geophys. Res. Atmos.]
  {10.1029/2017JD027931}, 123, 6864

\bibitem[\protect\citeauthoryear{Abbasi et~al.}{Abbasi
  et~al.}{2018b}]{Abbasi:2018nun}
Abbasi R.~U.,  et~al., 2018b, \mn@doi [Astrophys. J.]
  {10.3847/1538-4357/aabad7}, 858, 76

\bibitem[\protect\citeauthoryear{Abbasi et~al.}{Abbasi
  et~al.}{2019a}]{Abbasi:2018ywn}
Abbasi R.~U.,  et~al., 2019a, \mn@doi [Astropart. Phys.]
  {10.1016/j.astropartphys.2019.03.003}, 110, 8

\bibitem[\protect\citeauthoryear{Abbasi et~al.}{Abbasi
  et~al.}{2019b}]{Abbasi:2018wlq}
Abbasi R.~U.,  et~al., 2019b, \mn@doi [Phys. Rev.]
  {10.1103/PhysRevD.99.022002}, D99, 022002

\bibitem[\protect\citeauthoryear{Abraham et~al.}{Abraham
  et~al.}{2007}]{Abraham:2006ar}
Abraham J.,  et~al., 2007, \mn@doi [Astropart. Phys.]
  {10.1016/j.astropartphys.2006.10.004}, 27, 155

\bibitem[\protect\citeauthoryear{Abraham et~al.}{Abraham
  et~al.}{2008a}]{Aglietta:2007yx}
Abraham J.,  et~al., 2008a, \mn@doi [Astropart. Phys.]
  {10.1016/j.astropartphys.2008.01.003}, 29, 243

\bibitem[\protect\citeauthoryear{Abraham et~al.}{Abraham
  et~al.}{2008b}]{Abraham:2007rj}
Abraham J.,  et~al., 2008b, \mn@doi [Phys. Rev. Lett.]
  {10.1103/PhysRevLett.100.211101}, 100, 211101

\bibitem[\protect\citeauthoryear{{Abu-Zayyad, T. and others}}{{Abu-Zayyad, T.
  and others}}{2014}]{2014arXiv1403.0644T}
{Abu-Zayyad, T. and others} 2014, arXiv e-prints, \href
  {https://ui.adsabs.harvard.edu/abs/2014arXiv1403.0644T} {p. arXiv:1403.0644}

\bibitem[\protect\citeauthoryear{Abu-Zayyad et~al.}{Abu-Zayyad
  et~al.}{2012}]{AbuZayyad:2012hv}
Abu-Zayyad T.,  et~al., 2012, \mn@doi [Astrophys. J.]
  {10.1088/0004-637X/757/1/26}, 757, 26

\bibitem[\protect\citeauthoryear{Abu-Zayyad et~al.}{Abu-Zayyad
  et~al.}{2013a}]{AbuZayyad:2012ru}
Abu-Zayyad T.,  et~al., 2013a, \mn@doi [Astrophys. J.]
  {10.1088/2041-8205/768/1/L1}, 768, L1

\bibitem[\protect\citeauthoryear{Abu-Zayyad et~al.}{Abu-Zayyad
  et~al.}{2013b}]{Abu-Zayyad:2013dii}
Abu-Zayyad T.,  et~al., 2013b, \mn@doi [Phys. Rev.]
  {10.1103/PhysRevD.88.112005}, D88, 112005

\bibitem[\protect\citeauthoryear{Abu-Zayyad et~al.}{Abu-Zayyad
  et~al.}{2013c}]{AbuZayyad:2012kk}
Abu-Zayyad T.,  et~al., 2013c, \mn@doi [Nucl. Instrum. Meth.]
  {10.1016/j.nima.2012.05.079}, A689, 87

\bibitem[\protect\citeauthoryear{Agostinelli et~al.}{Agostinelli
  et~al.}{2003}]{Agostinelli:2002hh}
Agostinelli S.,  et~al., 2003, \mn@doi [Nucl. Instrum. Meth.]
  {10.1016/S0168-9002(03)01368-8}, A506, 250

\bibitem[\protect\citeauthoryear{{Alcantara}, {Anchordoqui}  \&
  {Soriano}}{{Alcantara} et~al.}{2019}]{2019arXiv190305429A}
{Alcantara} E.,  {Anchordoqui} L.~A.,   {Soriano} J.~F.,  2019, arXiv e-prints,
  \href {https://ui.adsabs.harvard.edu/abs/2019arXiv190305429A} {p.
  arXiv:1903.05429}

\bibitem[\protect\citeauthoryear{Ave, Hinton, Vazquez, Watson  \& Zas}{Ave
  et~al.}{2000}]{Ave:2000nd}
Ave M.,  Hinton J.~A.,  Vazquez R.~A.,  Watson A.~A.,   Zas E.,  2000, \mn@doi
  [Phys. Rev. Lett.] {10.1103/PhysRevLett.85.2244}, 85, 2244

\bibitem[\protect\citeauthoryear{Berezinsky, Kachelriess  \&
  Vilenkin}{Berezinsky et~al.}{1997}]{Berezinsky:1997hy}
Berezinsky V.,  Kachelriess M.,   Vilenkin A.,  1997, \mn@doi [Phys. Rev.
  Lett.] {10.1103/PhysRevLett.79.4302}, 79, 4302

\bibitem[\protect\citeauthoryear{Berezinsky, Blasi  \& Vilenkin}{Berezinsky
  et~al.}{1998}]{Berezinsky:1998ft}
Berezinsky V.,  Blasi P.,   Vilenkin A.,  1998, \mn@doi [Phys. Rev.]
  {10.1103/PhysRevD.58.103515}, D58, 103515

\bibitem[\protect\citeauthoryear{Bhattacharjee \& Sigl}{Bhattacharjee \&
  Sigl}{2000}]{Bhattacharjee:1998qc}
Bhattacharjee P.,  Sigl G.,  2000, \mn@doi [Phys. Rept.]
  {10.1016/S0370-1573(99)00101-5}, 327, 109

\bibitem[\protect\citeauthoryear{Bleve}{Bleve}{2016}]{Bleve:2015nut}
Bleve C.,  2016, PoS, ICRC2015, 1103

\bibitem[\protect\citeauthoryear{Blumenthal}{Blumenthal}{1970}]{Blumenthal:1970nn}
Blumenthal G.~R.,  1970, \mn@doi [Phys. Rev.] {10.1103/PhysRevD.1.1596}, D1,
  1596

\bibitem[\protect\citeauthoryear{Brun \& Rademakers}{Brun \&
  Rademakers}{1997}]{Brun:1997pa}
Brun R.,  Rademakers F.,  1997, \mn@doi [Nucl. Instrum. Meth.]
  {10.1016/S0168-9002(97)00048-X}, A389, 81

\bibitem[\protect\citeauthoryear{Cassiday et~al.}{Cassiday
  et~al.}{1990}]{Cassiday:1989yw}
Cassiday G.~L.,  et~al., 1990, \mn@doi [Nucl. Phys. Proc. Suppl.]
  {10.1016/0920-5632(90)90434-V}, 14A, 291

\bibitem[\protect\citeauthoryear{Coleman \& Glashow}{Coleman \&
  Glashow}{1999}]{Coleman:1998ti}
Coleman S.~R.,  Glashow S.~L.,  1999, \mn@doi [Phys. Rev.]
  {10.1103/PhysRevD.59.116008}, D59, 116008

\bibitem[\protect\citeauthoryear{Cummins \& Murphy}{Cummins \&
  Murphy}{2009}]{NLDN1}
Cummins K.,  Murphy M.~J.,  2009, IEEE Trans., 51, 499

\bibitem[\protect\citeauthoryear{Deligny, Kawata  \& Tinyakov}{Deligny
  et~al.}{2017}]{Deligny:2017wbx}
Deligny O.,  Kawata K.,   Tinyakov P.,  2017, \mn@doi [PTEP]
  {10.1093/ptep/ptx043}, 2017, 12A104

\bibitem[\protect\citeauthoryear{Fairbairn, Rashba  \& Troitsky}{Fairbairn
  et~al.}{2011}]{Fairbairn:2009zi}
Fairbairn M.,  Rashba T.,   Troitsky S.~V.,  2011, \mn@doi [Phys. Rev.]
  {10.1103/PhysRevD.84.125019}, D84, 125019

\bibitem[\protect\citeauthoryear{Feldman \& Cousins}{Feldman \&
  Cousins}{1998}]{Feldman:1997qc}
Feldman G.~J.,  Cousins R.~D.,  1998, \mn@doi [Phys. Rev.]
  {10.1103/PhysRevD.57.3873}, D57, 3873

\bibitem[\protect\citeauthoryear{Ferrari, Sala, Fasso  \& Ranft}{Ferrari
  et~al.}{2005}]{Ferrari:2005zk}
Ferrari A.,  Sala P.~R.,  Fasso A.,   Ranft J.,  2005, {preprint}, pp
  {CERN--2005--010}

\bibitem[\protect\citeauthoryear{Freund \& Schapire}{Freund \&
  Schapire}{1997}]{Freund1997119}
Freund Y.,  Schapire R.,  1997, \mn@doi [Journal of Computer and System
  Sciences] {10.1006/jcss.1997.1504}, 55, 119

\bibitem[\protect\citeauthoryear{Galaverni \& Sigl}{Galaverni \&
  Sigl}{2008}]{Galaverni:2007tq}
Galaverni M.,  Sigl G.,  2008, \mn@doi [Phys. Rev. Lett.]
  {10.1103/PhysRevLett.100.021102}, 100, 021102

\bibitem[\protect\citeauthoryear{Gelmini, Kalashev  \& Semikoz}{Gelmini
  et~al.}{2008}]{Gelmini:2005wu}
Gelmini G.,  Kalashev O.~E.,   Semikoz D.~V.,  2008, \mn@doi [J. Exp. Theor.
  Phys.] {10.1134/S106377610806006X}, 106, 1061

\bibitem[\protect\citeauthoryear{Glushkov, Gorbunov, Makarov, Pravdin, Rubtsov,
  Sleptsov  \& Troitsky}{Glushkov et~al.}{2007}]{Glushkov:2007ss}
Glushkov A.~V.,  Gorbunov D.~S.,  Makarov I.~T.,  Pravdin M.~I.,  Rubtsov
  G.~I.,  Sleptsov I.~E.,   Troitsky S.~V.,  2007, \mn@doi [JETP Lett.]
  {10.1134/S0021364007030010}, 85, 131

\bibitem[\protect\citeauthoryear{Glushkov, Makarov, Pravdin, Sleptsov,
  Gorbunov, Rubtsov  \& Troitsky}{Glushkov et~al.}{2010}]{Glushkov:2009tn}
Glushkov A.~V.,  Makarov I.~T.,  Pravdin M.~I.,  Sleptsov I.~E.,  Gorbunov
  D.~S.,  Rubtsov G.~I.,   Troitsky S.~V.,  2010, \mn@doi [Phys. Rev.]
  {10.1103/PhysRevD.82.041101}, D82, 041101

\bibitem[\protect\citeauthoryear{Gorbunov, Tinyakov, Tkachev  \&
  Troitsky}{Gorbunov et~al.}{2004}]{Gorbunov:2004bs}
Gorbunov D.~S.,  Tinyakov P.~G.,  Tkachev I.~I.,   Troitsky S.~V.,  2004,
  \mn@doi [JETP Lett.] {10.1134/1.1808838}, 80, 145

\bibitem[\protect\citeauthoryear{Gorski, Hivon, Banday, Wandelt, Hansen,
  Reinecke  \& Bartelman}{Gorski et~al.}{2005}]{Gorski:2004by}
Gorski K.~M.,  Hivon E.,  Banday A.~J.,  Wandelt B.~D.,  Hansen F.~K.,
  Reinecke M.,   Bartelman M.,  2005, \mn@doi [Astrophys. J.] {10.1086/427976},
  622, 759

\bibitem[\protect\citeauthoryear{Greisen}{Greisen}{1966}]{Greisen:1966jv}
Greisen K.,  1966, \mn@doi [Phys. Rev. Lett.] {10.1103/PhysRevLett.16.748}, 16,
  748

\bibitem[\protect\citeauthoryear{Heck, Knapp, Capdevielle, Schatz  \&
  Thouw}{Heck et~al.}{1998}]{Heck:1998vt}
Heck D.,  Knapp J.,  Capdevielle J.~N.,  Schatz G.,   Thouw T.,  1998,
  {preprint}, pp {FZKA--6019}

\bibitem[\protect\citeauthoryear{Hocker et~al.}{Hocker
  et~al.}{2007}]{Hocker:2007ht}
Hocker A.,  et~al., 2007, PoS, ACAT, 040

\bibitem[\protect\citeauthoryear{Homola, Gora, Heck, Klages, Pekala, Risse,
  Wilczynska  \& Wilczynski}{Homola et~al.}{2005}]{Homola:2003ru}
Homola P.,  Gora D.,  Heck D.,  Klages H.,  Pekala J.,  Risse M.,  Wilczynska
  B.,   Wilczynski H.,  2005, \mn@doi [Comput. Phys. Commun.]
  {10.1016/j.cpc.2005.07.001}, 173, 71

\bibitem[\protect\citeauthoryear{Hooper, Taylor  \& Sarkar}{Hooper
  et~al.}{2011}]{Hooper:2010ze}
Hooper D.,  Taylor A.~M.,   Sarkar S.,  2011, \mn@doi [Astropart. Phys.]
  {10.1016/j.astropartphys.2010.09.002}, 34, 340

\bibitem[\protect\citeauthoryear{Kachelriess, Kalashev  \&
  Kuznetsov}{Kachelriess et~al.}{2018}]{Kachelriess:2018rty}
Kachelriess M.,  Kalashev O.~E.,   Kuznetsov M.~{\relax Yu}.,  2018, \mn@doi
  [Phys. Rev.] {10.1103/PhysRevD.98.083016}, D98, 083016

\bibitem[\protect\citeauthoryear{Kalashev \& Kuznetsov}{Kalashev \&
  Kuznetsov}{2016}]{Kalashev:2016cre}
Kalashev O.~E.,  Kuznetsov M.~{\relax Yu}.,  2016, \mn@doi [Phys. Rev.]
  {10.1103/PhysRevD.94.063535}, D94, 063535

\bibitem[\protect\citeauthoryear{Kalashev \& Kuznetsov}{Kalashev \&
  Kuznetsov}{2017}]{Kalashev:2017ijd}
Kalashev O.~E.,  Kuznetsov M.~Y.,  2017, \mn@doi [JETP Lett.]
  {10.1134/S0021364017140016}, 106, 73

\bibitem[\protect\citeauthoryear{Kuzmin \& Rubakov}{Kuzmin \&
  Rubakov}{1998}]{Kuzmin:1997jua}
Kuzmin V.~A.,  Rubakov V.~A.,  1998, Phys. Atom. Nucl., 61, 1028

\bibitem[\protect\citeauthoryear{Kuznetsov}{Kuznetsov}{2017}]{Kuznetsov:2016fjt}
Kuznetsov M.~{\relax Yu}.,  2017, \mn@doi [JETP Lett.]
  {10.1134/S0021364017090028}, 105, 561

\bibitem[\protect\citeauthoryear{Maccione, Liberati  \& Sigl}{Maccione
  et~al.}{2010}]{Maccione:2010sv}
Maccione L.,  Liberati S.,   Sigl G.,  2010, \mn@doi [Phys. Rev. Lett.]
  {10.1103/PhysRevLett.105.021101}, 105, 021101

\bibitem[\protect\citeauthoryear{Matthews}{Matthews}{2018}]{Matthews:2017waf}
Matthews J.,  2018, \mn@doi [PoS] {10.22323/1.301.1096}, ICRC2017, 1096

\bibitem[\protect\citeauthoryear{Miller}{Miller}{1981}]{Miller:1991}
Miller R. G.~J.,  1981, {Simultaneous Statistical Inference}.
Springer-Verlag, New York, \mn@doi{10.1007/978-1-4613-8122-8}

\bibitem[\protect\citeauthoryear{NLDN}{NLDN}{}]{NLDNurl}
NLDN, {http://www.vaisala.com/en/products/
  thunderstormandlightningdetectionsystems/Pages/NLDN.aspx}

\bibitem[\protect\citeauthoryear{Nag et~al.}{Nag et~al.}{2011}]{NLDN2}
Nag A.,  et~al., 2011, J. Geophys. Res., 116, D02123

\bibitem[\protect\citeauthoryear{Nelson, Hirayama  \& Rogers}{Nelson
  et~al.}{1985}]{Nelson:1985ec}
Nelson W.~R.,  Hirayama H.,   Rogers D. W.~O.,  1985, {preprint}, pp
  {SLAC--0265}

\bibitem[\protect\citeauthoryear{Ostapchenko}{Ostapchenko}{2006}]{Ostapchenko:2004ss}
Ostapchenko S.,  2006, \mn@doi [Nucl. Phys. Proc. Suppl.]
  {10.1016/j.nuclphysbps.2005.07.026}, 151, 143

\bibitem[\protect\citeauthoryear{Raffelt \& Stodolsky}{Raffelt \&
  Stodolsky}{1988}]{Raffelt:1987im}
Raffelt G.,  Stodolsky L.,  1988, \mn@doi [Phys. Rev.]
  {10.1103/PhysRevD.37.1237}, D37, 1237

\bibitem[\protect\citeauthoryear{Risse \& Homola}{Risse \&
  Homola}{2007}]{Risse:2007sd}
Risse M.,  Homola P.,  2007, \mn@doi [Mod. Phys. Lett.]
  {10.1142/S0217732307022864}, A22, 749

\bibitem[\protect\citeauthoryear{Risse, Homola, Engel, Gora, Heck, Pekala,
  Wilczynska  \& Wilczynski}{Risse et~al.}{2005}]{Risse:2005jr}
Risse M.,  Homola P.,  Engel R.,  Gora D.,  Heck D.,  Pekala J.,  Wilczynska
  B.,   Wilczynski H.,  2005, \mn@doi [Phys. Rev. Lett.]
  {10.1103/PhysRevLett.95.171102}, 95, 171102

\bibitem[\protect\citeauthoryear{Ros, Supanitsky, Medina-Tanco, del Peral  \&
  Rodríguez-Frías}{Ros et~al.}{2013}]{Ros:2013lxa}
Ros G.,  Supanitsky A.~D.,  Medina-Tanco G.~A.,  del Peral L.,
  Rodríguez-Frías M.~D.,  2013, \mn@doi [Astropart. Phys.]
  {10.1016/j.astropartphys.2013.05.014}, 47, 10

\bibitem[\protect\citeauthoryear{Rubtsov \& Troitsky}{Rubtsov \&
  Troitsky}{2015}]{Rubtsov:2015wba}
Rubtsov G.~I.,  Troitsky S.~V.,  2015, \mn@doi [J. Phys. Conf. Ser.]
  {10.1088/1742-6596/608/1/012067}, 608, 012067

\bibitem[\protect\citeauthoryear{Rubtsov et~al.}{Rubtsov
  et~al.}{2006}]{Rubtsov:2006tt}
Rubtsov G.~I.,  et~al., 2006, \mn@doi [Phys. Rev.]
  {10.1103/PhysRevD.73.063009}, D73, 063009

\bibitem[\protect\citeauthoryear{Rubtsov, Satunin  \& Sibiryakov}{Rubtsov
  et~al.}{2012}]{Rubtsov:2012kb}
Rubtsov G.,  Satunin P.,   Sibiryakov S.,  2012, \mn@doi [Phys. Rev.]
  {10.1103/PhysRevD.86.085012}, D86, 085012

\bibitem[\protect\citeauthoryear{Rubtsov, Satunin  \& Sibiryakov}{Rubtsov
  et~al.}{2014}]{Rubtsov:2013wwa}
Rubtsov G.,  Satunin P.,   Sibiryakov S.,  2014, \mn@doi [Phys. Rev.]
  {10.1103/PhysRevD.89.123011}, D89, 123011

\bibitem[\protect\citeauthoryear{Shinozaki et~al.}{Shinozaki
  et~al.}{2002}]{Shinozaki:2002ve}
Shinozaki K.,  et~al., 2002, \mn@doi [Astrophys. J.] {10.1086/341288}, 571,
  L117

\bibitem[\protect\citeauthoryear{Sommers}{Sommers}{2001}]{Sommers:2000us}
Sommers P.,  2001, \mn@doi [Astropart. Phys.] {10.1016/S0927-6505(00)00130-4},
  14, 271

\bibitem[\protect\citeauthoryear{Stokes, Cady, Ivanov, Matthews  \&
  Thomson}{Stokes et~al.}{2012}]{Stokes:2011wf}
Stokes B.~T.,  Cady R.,  Ivanov D.,  Matthews J.~N.,   Thomson G.~B.,  2012,
  \mn@doi [Astropart. Phys.] {10.1016/j.astropartphys.2012.03.004}, 35, 759

\bibitem[\protect\citeauthoryear{Teshima et~al.}{Teshima
  et~al.}{1986}]{Teshima:1986rq}
Teshima M.,  et~al., 1986, \mn@doi [J. Phys.] {10.1088/0305-4616/12/10/017},
  G12, 1097

\bibitem[\protect\citeauthoryear{Tokuno et~al.}{Tokuno
  et~al.}{2012}]{Tokuno:2012mi}
Tokuno H.,  et~al., 2012, \mn@doi [Nucl. Instrum. Meth.]
  {10.1016/j.nima.2012.02.044}, A676, 54

\bibitem[\protect\citeauthoryear{Zatsepin \& Kuzmin}{Zatsepin \&
  Kuzmin}{1966}]{Zatsepin:1966jv}
Zatsepin G.~T.,  Kuzmin V.~A.,  1966, JETP Lett., 4, 78

\makeatother
\end{thebibliography}

\vspace*{0.5cm}
\noindent
Affiliations:\\
{\it
$^{1}$ High Energy Astrophysics Institute and Department of Physics and Astronomy, University of Utah, Salt Lake City, Utah, USA \\
$^{2}$ The Graduate School of Science and Engineering, Saitama University, Saitama, Saitama, Japan \\
$^{3}$ Graduate School of Science and Engineering, Tokyo Institute of Technology, Meguro, Tokyo, Japan \\
$^{4}$ Department of Physics and The Research Institute of Natural Science, Hanyang University, Seongdong-gu, Seoul, Korea \\
$^{5}$ Department of Physics, Tokyo University of Science, Noda, Chiba, Japan \\
$^{6}$ Department of Physics, Kindai University, Higashi Osaka, Osaka, Japan \\
$^{7}$ Service de Physique Theorique, Universite Libre de Bruxelles, Brussels, Belgium \\
$^{8}$ Hakubi Center for Advanced Research, Kyoto University, Sakyo-ku, Kyoto, Japan \\
$^{9}$ Graduate School of Science, Kyoto University, Sakyo-ku, Kyoto, Japan \\
$^{10}$ Graduate School of Science, Osaka City University, Osaka, Osaka, Japan \\
$^{11}$ Institute for Cosmic Ray Research, University of Tokyo, Kashiwa, Chiba, Japan \\
$^{12}$ Kavli Institute for the Physics and Mathematics of the Universe (WPI), Todai Institutes for Advanced Study, University of Tokyo, Kashiwa, Chiba, Japan \\
$^{13}$ Information Engineering Graduate School of Science and Technology, Shinshu University, Nagano, Nagano, Japan \\
$^{14}$ Faculty of Engineering, Kanagawa University, Yokohama, Kanagawa, Japan \\
$^{15}$ Interdisciplinary Graduate School of Medicine and Engineering, University of Yamanashi, Kofu, Yamanashi, Japan \\
$^{16}$ Astrophysical Big Bang Laboratory, RIKEN, Wako, Saitama, Japan \\
$^{17}$ Department of Physics, Sungkyunkwan University, Jang-an-gu, Suwon, Korea \\
$^{18}$ Department of Physics, Tokyo City University, Setagaya-ku, Tokyo, Japan \\
$^{19}$ Institute for Nuclear Research of the Russian Academy of Sciences, Moscow, Russia \\
$^{20}$ Advanced Research Institute for Science and Engineering, Waseda University, Shinjuku-ku, Tokyo, Japan \\
$^{21}$ Department of Physics, Chiba University, Chiba, Chiba, Japan \\
$^{22}$ Department of Physics, School of Natural Sciences, Ulsan National Institute of Science and Technology, UNIST-gil, Ulsan, Korea \\
$^{23}$ Department of Physics, Yonsei University, Seodaemun-gu, Seoul, Korea \\
$^{24}$ Academic Assembly School of Science and Technology Institute of Engineering, Shinshu University, Nagano, Nagano, Japan \\
$^{25}$ Faculty of Science, Kochi University, Kochi, Kochi, Japan \\
$^{26}$ Nambu Yoichiro Institute of Theoretical and Experimental Physics, Osaka City University, Osaka, Osaka, Japan \\
$^{27}$ Department of Physical Sciences, Ritsumeikan University, Kusatsu, Shiga, Japan \\
$^{28}$ Sternberg Astronomical Institute, Moscow M.V. Lomonosov State University, Moscow, Russia \\
$^{29}$ Department of Physics and Astronomy, Rutgers University - The State University of New Jersey, Piscataway, New Jersey, USA \\
$^{30}$ Earthquake Research Institute, University of Tokyo, Bunkyo-ku, Tokyo, Japan \\
$^{31}$ Department of Engineering Science, Faculty of Engineering, Osaka Electro-Communication University, Neyagawa-shi, Osaka, Japan \\
$^{32}$ Graduate School of Information Sciences, Hiroshima City University, Hiroshima, Hiroshima, Japan \\
$^{33}$ Institute of Particle and Nuclear Studies, KEK, Tsukuba, Ibaraki, Japan \\
$^{34}$ National Institute of Radiological Science, Chiba, Chiba, Japan \\
$^{35}$ CEICO, Institute of Physics, Czech Academy of Sciences, Prague, Czech Republic \\
$^{36}$ Department of Physics and Institute for the Early Universe, Ewha Womans University, Seodaaemun-gu, Seoul, Korea \\
$^{37}$ Department of Physics, Ehime University, Matsuyama, Ehime, Japan \\
}

\bsp	
\label{lastpage}
\end{document}